\documentclass[twocolumn]{aastex63}
\usepackage{comment}
 \usepackage{hyperref}
\usepackage{amsmath}
 \usepackage{xfrac}
\usepackage{color}
 \usepackage{hyperref}
\usepackage{datetime2}
\usepackage{xcolor}

\newcommand\hd{HD\,163466}
\newcommand{\Stromgren}{Str\"omgren}
 \received{\today\ \  \DTMcurrenttime}

\shorttitle{Warm Ionized Medium}
 \shortauthors{Kulkarni, Beichman \&\ Ressler}

\newcommand{\mum}{$\mu$m}

\begin{document}

\title{Mid-infrared fine structure lines from the Galactic warm
ionized medium}

\correspondingauthor{S.\ R.\ Kulkarni}
 \email{srk@astro.caltech.edu}

\author[0000-0001-5390-8563]{S.\ R.\ Kulkarni}
 \affiliation{Department of Astronomy, Cornell University, Ithaca, NY\,14853}
\author[0000-0002-5627-5471]{Charles  Beichman}
 \affiliation{NASA Exoplanet Science Institute, Jet Propulsion
 Laboratory, California Institute of Technology, 1200 East California
 Blvd, Pasadena, CA 91125, USA}
\author[0000-0001-5644-8830]{Michael E.\ Ressler}
 \affiliation{Jet Propulsion Laboratory,
California Institute of Technology, 4800 Oak Grove Drive, Pasadena,
CA 91109, USA}

\begin{abstract}
 The Warm Ionized Medium (WIM)  hosts most of the ionized gas in the
 Galaxy and occupies perhaps a quarter of the volume of the Galactic
 disk. Decoding the spectrum of the Galactic diffuse ionizing field is
 of fundamental interest. This can be done via direct measurements
 of ionization fractions of various elements. 
  Based on current physical models for the WIM we predicted
 that mid-IR fine structure lines of Ne, Ar and S would be within the grasp of 
 the Mid-Infrared Imager-Medium Resolution Spectrometer (MIRI-MRS), an Integral Field Unit (IFU) spectrograph,
 aboard the James Webb Space Telescope (JWST).  Motivated thus we
 analyzed a pair of commissioning data sets and detected [NeII]\,12.81\,$\mu$m,
 [SIII]\,18.71\,$\mu$m and possibly [SIV]\,10.51\,$\mu$m.  
 The inferred emission measure for these detections is about $10\,{\rm cm^{-6}\,pc}$, typical of the WIM. These detections are broadly consistent
 with expectations of physical models for the WIM.  The current detections
 are limited by uncorrected fringing (and to a lesser extent by
 baseline variations). In due course, we
 expect, as with other IFUs, the calibration pipeline to deliver
 photon-noise-limited spectra.  The detections reported here
 bode well for the study of the WIM.
 Along most lines-of-sight hour-long MIRI-MRS observations
 should detect line emission from the
 WIM. When combined with optical observations by modern IFUs with
 high spectral resolution on large ground-based telescopes, the
 ionization fraction  and temperature of neon and sulfur can be
 robustly inferred. Separately, the ionization of helium in the WIM can be probed
 by NIRspec.
 Finally, joint JWST and optical IFU studies  will 
 open up 
 a new cottage industry of studying the WIM on arcsecond scales.\\
\end{abstract}

\section{Introduction}

A major challenge in understanding the WIM
is the energetics and propagation of Lyman continuum.
The inferred ionizing
power is high, requiring a sixth or so of the Lyman continuum
($\lambda<912\,$\AA, hereafter the Extreme Ultraviolet or EUV)
output of the Galactic OB stars (see \citealt{hdb+09} for review; \citealt{R84}). Separately, in order
to explain the filling factor, the EUV photons, although originating
in the Galactic plane, have to diffuse to nooks and crannies in the
Galactic disk. The spectrum of the ionizing photons is significantly modified
as they propagate away from star-forming regions.
The relative abundance of ions in various ionization
states allow us to directly probe the spectrum
of the diffuse EUV radiation field.

The traditional diagnostics of the WIM have been optical recombination
lines of hydrogen and helium and optical nebular lines of O~I, N~I,
O~II, N~II, S~II and O~III.  An illustrative example of the richness
of the optical lines can be found in \cite{R00}.  The Wisconsin
H$\alpha$ Mapper (WHAM; \citealt{T97,rth+98}) was and continues to
be the primary workhorse for optical studies of the WIM. It undertook
a full-sky H$\alpha$ survey as well as several large-area imagery
in [SII] and [NII].  The line ratios (e.g., [SII]/H$\alpha$,
[NII]/[SII]) for the WIM are very different from those seen towards
bright H~II regions.

We summarize the key findings that have emerged from the optical
studies.  A number of lines of evidence argue for a WIM temperature
of 8,000\,K to 10,000\,K, significantly higher than those of H~II
regions (5,000\,K to 7,000\,K; depending on metalicity).  Three
deep observations of [OI]\,$\lambda$6300\,\AA\ constrain the neutral
fraction of hydrogen, thanks to the strong charge exchange, to 10\%
\citep{hrh02}.  A single deep observation in [NI]\,$\lambda$5100\,\AA\
constrains the neutral fraction of nitrogen, $n_{\rm N^0}/n_{\rm
N}\lesssim 0.05$ \citep{rrs77}.  The consensus  view is that
the hydrogen in the WIM is partially ionized, $x_{\rm
H^+}=n({\rm H^+})/n({\rm H}) \approx 0.9$.
The few observations in the [OIII] provide an upper limit to the temperature of the WIM,
$\lesssim 10^4\,$K \citep{R85,mr05}. The WIM
was investigated in [OII] with a novel 
spectrometer \citep{mrr+06} but no long-term program was undertaken.

The ``ionization parameter", $U$, the ratio of the number density
of ionizing photons to the number density of hydrogen atoms, is key
to distinguishing classical H~II regions from the WIM.  The Lyman
continuum optical depth to the \Stromgren\ surface, $\tau_0\propto
U^{1/3}$.  Classical H~II regions with $\tau_0\gg 1$ have sharply
defined \Stromgren\ surfaces with gas within the sphere fully ionized.
The WIM optical line ratios discussed above have been traditionally
modeled by low values of $U$ \citep{M86,dm94,shr+00}.  Observations
of an optical helium recombination line \citep{rt95} and extensive
and deep radio recombination line studies \citep{hkl+96} suggest
that helium is weakly ionized, $x_{\rm He^+}<0.3$ (optical) or  $<0.13$
(radio).  This suggests that the diffuse EUV field is relatively
``soft".



Through  the  fine
structure lines of Ne~II, Ne~III, Ar~II, Ar~III, S~III and S~IV,
and recombination lines of hydrogen and helium
the mid-IR offer new diagnostics of the WIM. The recently launched James Webb Space
Telescope (JWST) carries two spectrometers: the Near Infrared
Spectrograph (NIRSpec; \citealt{bal+22,jfo+22}) covering the
wavelength range 1--5\,$\mu$m while the Mid-Infrared Instrument-Medium
Resolution Spectrograph (MIRI-MRS; \citealt{wpg+15,lab+21}) covers
5--29\,$\mu$m.  JWST, with its narrow spectroscopic field-of-view
(FoV, $\Omega$ ranging from ten to fifty arcsec$^2$) is not an 
obvious facility of choice to study diffuse emission from Galactic
WIM.  However, at turns out,  this small FoV is compensated
by the large collecting area of JWST, efficient integral field unit (IFU) spectrometers with
spectral resolution of few thousand \citep{wpg+15} and a thousand spaxels and  detectors
with negligible dark current \citep{rsf+15}.

The paper is organized as follows.  In \S\ref{sec:FSL} we introduce
the mid-IR fine structure lines and summarize their potential
diagnostic value.   In \S\ref{sec:MRS_Detectability} we investigate
the detectability of mid-IR fine structure lines with MIRI-MRS.
Buoyed by the conclusions of our study we analyzed MIRI-MRS
commissioning data sets.  In \S\ref{sec:MRS_Detection} we present
secure detections of Ne~II and S~III and possibly S~IV detection
from the WIM. 
In \S\ref{sec:Inference} we find the detections are consistent with
the low-$U$ photoionization models invoked for the WIM.
We conclude
in \S\ref{sec:Conclusions} by first noting that hour-long MIRI-MRS
observations will, for many lines-of-sight, detect of  [NeII], [ArII]
and [SIII] from the WIM. Thus, there will be a steady growth of
measurements of the WIM ionization fractions.  We  discuss the
detectability of ionized helium with NIRSpec.  We end by noting the
great returns that would be made possible by joint studies undertaken
with JWST \&\ ground-based high spectral-resolution IFU spectrographs.

Unless otherwise mentioned, all basic formulae, collisional and
recombination coefficients are from \cite{D11} and the atomic data
(A-coefficients, wavelengths) from 
NIST\footnote{\url{https://www.nist.gov/pml/atomic-spectra-database}\label{fn:NIST}}.

\section{The Mid-IR Fine Structure Lines}
 \label{sec:FSL}

In the optical, 
line ratios have been used to infer the temperature and study
the state of ionization. For instance,
WHAM observations of variation
in  [NII]/[SII] is nicely explained by variation in $x_{\rm S^+}$
from 0.3 to 0.8 while variations in [NII]/H$\alpha$ are readily
explained by temperature variations, from 6,000\,K to 10,000\,K
\citep{hrt99}. Next, the low value of $x_{\rm S^{++}}$ (inferred from [SIII] observations) is suggestive of the
softening of the diffuse EUV field even at photon energies as low
as 23\,eV.

The mid-IR lines, unlike the optical lines, are not sensitive
to variations in temperature and furthermore suffer much less from extinction
(including reflection).
As will become clear from the
discussion below they are also well suited to probing
the ionization state.

Of the elements with more than one part million (ppm), Ne, Ar and S have mid-IR
lines that are accessible to JWST\footnote{Fine structure lines of highly ionized species such
as NeV and OIV are not of interest to WIM studies.}.
As can be seen from Table~\ref{tab:IP_abundance} the ionization
potentials of  Ne, Ar and S are well suited to probing the spectrum
of the diffuse EUV radiation field above the H~I and He~II edges.  
Below we develop the formulae for  intensities
of the mid-IR fine structure lines. The fine structure splitting of [NeII], [ArII]
and [SIV] results in two levels but 
[NeIII], [ArIII] and [SIII] have three levels. 
Note that [NeIII]\,30.01\,$\mu$m
and [SIII]\,33.48\,$\mu$m  lie outside the wavelength range of
MIRI-MRS and so are dropped from any further discussion.
The relevant atomic physics data can be found
in Appendix~\ref{sec:AtomicData}.  


\begin{deluxetable}{lllrr}[hbt]
\label{tab:IP_abundance}
\tablecaption{Ionization Potential}
\tablehead{
 \colhead{$X$} &
 \colhead{$Y_X$\,(ppm)} & 
 \colhead{I$\rightarrow$II} &
 \colhead{II$\rightarrow$III} &
 \colhead{III$\rightarrow$IV} 
}
\startdata
Ne & 93.3   & 21.6 & 41.0 & 63.4\\
S   & 14.5   & 10.4 & 23.3 & 34.8 \\
Ar  & 2.75   & 15.8  & 27.6 & 40.7\\
\hline
N  & 74.1    & 14.5 & 29.6 & 47.4
\enddata
\tablecomments{$X$ is element and $Y_{X}$ is the
abundance by number, relative to hydrogen, in
parts per million (ppm). Subsequent columns are 
ionization potential in eV. Nitrogen is included as
a point of comparison to argon. For reference, the
ionization potential of helium is 24.587\,eV.}
\end{deluxetable}

\subsection{ Neon}
 \label{sec:Neon}

The  [NeII]\,$12.813\,\mu$m fine structure line arises from the
spin-orbit splitting of the $^2{\rm P}^o$ ground term.   For
simplicity we assume that the electrons are provided only by
the ionization of hydrogen,
$n_e=n_{\rm H^+}$.  The rate of electron excitations per unit volume is 
$C=n_{\rm H^+}n_{\rm Ne^+}q_{lu}$ where
 \begin{equation*}
   q_{lu} = \frac{8.629\times 10^{-8}}{T_4^{1/2}}
    \frac{\Omega_{ul}}{g_l} {\rm e}^{-T_{u}/T}\,{\rm cm^3\,s^{-1}}
 \end{equation*} 
is the electron collisional coefficient, $g_l$ is the degeneracy
factor of the lower level; $\Omega_{ul}$ is the collisional strength
and  is given in Table~\ref{tab:Atomic_FSL}; and $T_{u}=E_{u}/k_B$
with  $E_{u}$ being the energy level difference between the upper
and the ground state.\footnote{
For the low density  of the WIM, given the values of A-coefficients
for the fine structure lines (see Table~\ref{tab:Atomic_FSL}), 
we can safely assume that, in the WIM,
almost all  atoms and ions are in the ground state.}
 The photon intensity
is  $\int Cdl /(4\pi)$, the integral along the line-of-sight,
 \begin{eqnarray}
  I_{\rm [NeII]} &=& 1.95\xi_{\rm Ne^+}T_4^{-0.424+0.002\,{\rm
  ln}T_4} {\rm e}^{-0.112/T_4}{\rm EM}\ R
   \label{eq:NeII}
 \end{eqnarray}
where $\xi_{\rm Ne^+}=x_{\rm Ne^+}/x_{\rm H^+}$ and ${\rm EM}=\int
n_{\rm H^+}^2dl$ is the emission measure carrying the unit of ${\rm
cm^{-6}\,pc}$ and $R$ stands for Rayleigh.\footnote{Recall that one
Rayleigh is $10^6/(4\pi)\,{\rm photon\, cm^{-2}\, s^{-1}\,ster^{-1}}$
which translates to $7.96\times 10^4\,{\rm photon\, cm^{-2}\,
s^{-1}\,ster^{-1}}$ or $1.87\times 10^{-6}\,{\rm photon\, cm^{-2}\,
s^{-1}\,arcsec^{-2}}$.} For reference, the EM from the WIM varies
from $1\,{\rm cm^{-6}\,pc}$ (Galactic poles) to $25\,{\rm cm^{-6}\,pc}$
(the ``brightest" WIM region; see \citealt{mrh06}). For the WIM,
we assume a fiducial temperature of 8,000\,K. At this temperature,
$I_{\rm Ne^+}= 1.86\,{\rm EM}\,R$.

For Ne~III, excitation from the ground level to the
first and second levels result in emission of 15.555\,$\mu$m photons.  The intensity
of the 15.555\,$\mu$m line is given by the sum of the excitations:
 \begin{multline*}
  I_{\rm [NeIII]} = \xi_{\rm Ne^{+2}}\Big[
 3.85T_4^{-0.432-0.0556\,{\rm ln}\,T_4}\,{\rm e}^{-0.0925/T_4}+ \\
1.03T_4^{-0.440 -0.053\,{\rm ln}\,T_4}\,{\rm e}^{-0.1324/T_4}\Big]\,{\rm
EM}\,R \label{eq:NeIII}
 \end{multline*}
with $\xi_{\rm Ne^{+2}}=x_{\rm Ne^{+2}}/x_{\rm H^+}$, and, as before,
the atomic data can be found in \S\ref{sec:AtomicData}.  At the
fiducial temperature, $I_{\rm [NeIII]}\approx 4.73\xi_{\rm
Ne^{+2}}\,{\rm EM}\,R$.

\subsection{Argon \&\ Sulfur}
 \label{sec:ArgonSulfur}

In a similar manner the intensities of fine structure lines of
argon and sulfur (\S\ref{sec:AtomicData}) can be computed:
 \begin{eqnarray*}
  I_{\rm [ArII]} &=& 0.72\xi_{\rm Ar^+} T_4^{-0.416-0.014{\rm
  ln}\,T_4}{\rm e}^{-0.206/T_4}\,{\rm EM} \,R\ ,\\ 
  I_{\rm [ArIII]}&&=\xi_{\rm Ar^{+2}}\Big[
  0.59T_4^{-0.469+0.002{\rm ln}\,T_4}{\rm e}^{-0.1601/T_4}\cr
  &&\ \ +0.15T_4^{-0.389-0.009{\rm ln}\,T_4}{\rm e}^{-0.2259/T_4}\Big]{\rm EM}\,R\ ,\\
  I_{\rm [SIII]}&=&30.4\xi_{\rm S^{+2}}T_4^{-0.671 - 0.033{\rm ln}\,T_4}{\rm e}^{-0.1199/T_4}\,{\rm EM}\,R\ , \\
  I_{\rm [SIV]}&=&16.5\xi_{\rm S^{+3}}T_4^{-0.512-0.076{\rm ln}\,T_4}{\rm e}^{-0.1369/T_4}\,
  {\rm EM}\,R\ .
 \end{eqnarray*}

\begin{deluxetable*}{llllllrrrr}
 \tablecaption{Detectability with MIRI-MRS}
 \label{tab:SNR}
\tablewidth{0pt}
\tablehead{
 \colhead{species} &
 \colhead{$\lambda$} & 
 \colhead{ChB} & 
 \colhead{$\mathcal{R}$} & 
  \colhead{$\theta^{\prime\prime}$} & 
 \colhead{$\eta$} &
 \colhead{$B_\nu$} &
  \colhead{$\sigma\,(R)$}  &
 \colhead{$S/{\rm EM}\,(R)$} &
 \colhead{$S/\sigma$}
 }
\startdata
{[}NeII]& 12.81 & 3A & 2880& 6.1& 0.11 & 30 & 0.17 & 
	$1.85\xi_{\rm Ne^+}$ & $10.9\xi_{\rm Ne^+}{\rm EM}$\\
{[}NeIII] &15.55 & 3B & 2560& 6.1& 0.11 & 60 & 0.25 &  
	$4.73\xi_{\rm Ne^+2}$ & $18.8\xi_{\rm Ne^{+2}}{\rm EM}$\\
{[}ArII] &6.98 & 1C & 3300& 3.7& 0.14 & 3  & 0.07 & 
	$0.91\xi_{\rm Ar^+}$ & $12.5\xi_{\rm Ar^+}{\rm EM}$\\
{[}ArIII] & 8.99 & 2B & 2850 & 4.6 & 0.13 & 5 & 0.08 & 
	$0.66\xi_{\rm Ar^{+2}}$ & $7.8\xi_{\rm Ar^{+2}}{\rm EM}$   \\
{[}ArIII]&  21.83 & 4B&  1700 & 7.8 & 0.02 & 300 & 1.27 & 
	$0.12\xi_{\rm Ar^{+2}}$ & $0.09\xi_{\rm Ar^{+2}}{\rm EM}$\\
{[}SIII] & 18.71 & 4A  & 1610& 7.8& 0.03& 160 & 0.78 & 
	$30.33\xi_{\rm S^{+2}}$ & $39.0\xi_{\rm S^{+2}}{\rm EM}$\\
{[}SIV] & 10.51 & 2C & 3000 & 4.6 & 0.134 & 20 & 0.16 & 
	$15.52\xi_{\rm S^{+3}}$ & $96.1\xi_{\rm S^{+3}}{\rm EM}$
\enddata
 \tablecomments{\small The wavelength, $\lambda$, is in microns.
 ``ChB" refers to Channel-band combination.  The spectral resolution,
 $\mathcal{R}=\lambda/\Delta\lambda$ and the side-length of the IFU
 field-of-view, $\theta$, are from Table~1 of \cite{wpg+15} while
 $\eta$ is from Table~5 of \citet{wpg+15}.  $B_\nu$ is a rough
 estimate\footnote{from
 \url{https://jwst-docs.stsci.edu/jwst-general-support/jwst-background-model}}
 of the  net background emission in MJy\,ster$^{-1}$. $\sigma$ is
 the Poisson uncertainty due to the background, assuming $t=3600\,s$
 of integration time, in a spectral channel carrying the unit of
 Rayleigh. In computing $S$, the signal strength in Rayleigh, we
 set $T=8,000\,$K, Note that $\xi_{\rm X}=x_{\rm X}/x_{\rm H^+}$
 where $x_{\rm X}$ is the ionization fraction of species $X$. EM
 is the emission measure along the line-of-sight (unit: cm$^{-6}$\,pc).
 The last column is the expected signal-to-noise ratio. }
\end{deluxetable*}

\subsection{Ionization Fraction}
 \label{sec:IonizationFraction}
 
In order to infer the ionization fraction of, for instance
[NeII], we need to know the emission measure.  The case~B H$\alpha$
photon intensity is given by
 \begin{equation*}
    I_{\rm H\alpha} = 0.361 T_4^{-0.942-0.031\,{\rm ln}\,T_4}\,{\rm
    EM}\,R \ .
  \label{eq:Halpha}
 \end{equation*}
The ratio
 \begin{equation*}
  \frac{\rm [NeII]}{\rm H\alpha} \equiv \frac{I_{\rm Ne^+}}{I_{\rm
  H\alpha}}= 5.4\xi_{\rm Ne^+}T_4^{0.518+0.033{\rm ln}\,T_4}
    {\rm e}^{-0.112/T_4}
 \end{equation*}
is a direct measure of $\xi_{\rm Ne^+}$. Notice, relative to similar
ratios involving forbidden optical lines, the ratio (above)  is a weak(er)
function of temperature.  The primary source of H$\alpha$
data is from WHAM which has a beam of one degree diameter while the
MIRI-MRS FoV is about ten arcsec$^2$ (see Table~\ref{tab:SNR}).
There is no reason to believe that the WIM is smooth on scales of
a degree. We will return to this important point in the concluding section.

\begin{deluxetable*}{lrrrr}[hbt]
\label{tab:logHD}
\tablecaption{Summary of log of data sets for\hd}
\tablehead{
\colhead{name}&
\colhead{$\alpha$\,(deg)} &
\colhead{$\delta$\,(deg)} &
\colhead{series} &
\colhead{$\tau$(s)}
}
\startdata
HD163466 & 268.1057 & 60.396 & \texttt{jw01050-o009\_t004} & 10,930\\
HD-BKG1  & 268.0731 & 60.418 & \texttt{jw01050-o008\_t011}  & 1,682 \\
HD-BKG2  & 268.0731 & 60.418 & \texttt{jw01050-o010\_t011}  & 1,682
\enddata
 \tablecomments{ The ``name" is our assigned name for the data set.
 The next two columns are J2000 right ascension and declination,
 followed by file name identifier  of Level-3 MIRI-MRS pipeline
 data sets.  Each data set has four files, one for each channel. For
 instance, for HD-BKG1, the data cube for the first channel is
 \texttt{jw01050-o010\_t011\_ch1-longshortmedium-\_s3d.fits} while
 that for the second channel is
 \texttt{jw01050-o010\_t011\_ch2-longshortmedium-\_s3d.fits} and
 so on.  The last column is the integration time.  The program ID
 (PID) for these data sets is 1050.}
\end{deluxetable*}

\section{Detectability with MIRI/Medium Resolution Spectrometer}
 \label{sec:MRS_Detectability}
 
The  Medium Resolution Spectrometer is an integral field unit (IFU)
spectrograph with a spectral resolution, $\mathcal{R}=\lambda/{\rm
FWHM}$ of 2000 to 3000  \citep{wpg+15,lab+21}; here FWHM is the
full-width at half-maximum of an unresolved line.  The instrument
is quite complex with four simultaneous channels (1--4) and three
selectable bands (A, B, C).  The full wavelength range,
4.87--28.82\,$\mu$m, is covered by successively going through the
three bands.  The entrance aperture of the IFU is approximately a
square of side $\theta$ which varies from  $3.7^{\prime\prime}$
(Channel 1) to $7.8^{\prime\prime}$ (Channel 4).  The
``photon-photoelectron conversion" efficiency (i.e., the net
throughput of the telescope, spectrometer optics and quantum
efficiency of the detectors), $\eta$, and $\mathcal{R}$ depend on
the channel-band combination.  The instrumental parameters are
summarized in Table~\ref{tab:SNR}.  We assume that the detector 
dark current is
negligible.

Our goal is to measure the mid-IR spectrum of the sky and we do so
by using MRS as a ``light bucket". The background intensity, $B_\nu$, is
usually quoted in units of MJy\,ster$^{-1}$  and arises from a combination of thermal emission from the  local zodiacal dust cloud and at wavelengths long-ward of $\sim$15 \mum, self-emission from the telescope. Between 5 and 15 \mum\ the background intensity\footnote{https://jwst-docs.stsci.edu/jwst-general-support\\ /jwst-background-model} increases from 0.5 to 50 MJy\,ster$^{-1}$.   Then the rate of
photo-electrons from the background in one spectral resolution
element, frequency width $\Delta\nu\approx \nu/\mathcal{R}$, is
$(B_\nu/h\mathcal{R})A\Omega\eta$ where $A=25.4\,{\rm m}^2$ is the
collecting area of JWST and $\Omega=\theta^2$.  The thermal line
width is a few  km\,s$^{-1}$. Including Galactic rotation,
the expected [NeII] line will be effectively confined to an
effective spectral channel (but the FWHM, depending on wavelength,
is spread between 2 and 4  pixels; see Figure~14 of \citealt{wpg+15}).
Because the zodiacal and/or telescope  emission is smooth both spatially across a few arcseconds and  across a few spectral channels, background subtraction to reveal a spectral line should be photon-noise limited once detector calibration issues are fully resolved.

The rate of photo-electrons due to line emission from the WIM is
$SA\Omega\eta$ where $S$ is the line photon intensity (${\rm
phot\,cm^{-2}\,ster^{-1}\,s^{-1}}$). The SNR of the line is then
 \begin{equation}
  {\rm
  SNR}=\frac{S}{\sqrt{B_\nu/h}}\Big(A\Omega\eta\mathcal{R}t\Big)^{1/2}
   \label{eq:SNR}
 \end{equation}
where $t$ is the integration time.
In Table~\ref{eq:SNR} we summarize the detectability by MRS. We
also list the specific channel-band in which these lines can
be observed.  As can be gathered from this Table, detections and
useful upper limits can be obtained for all species provided that the 
EM is greater than a few units.

\begin{figure}[htbp]            
 \plotone{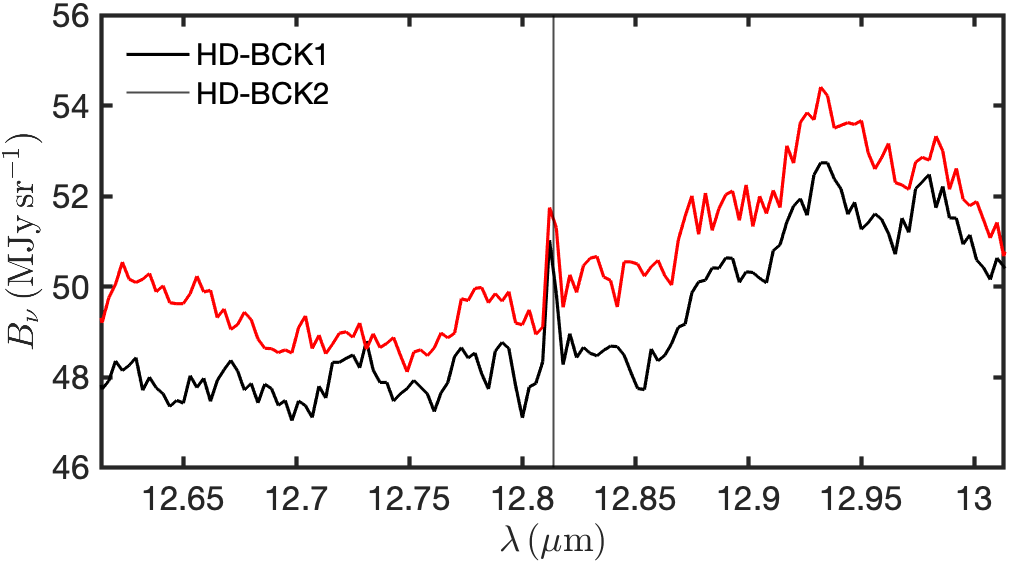}
  \caption{\small The spectra of the sky, in the vicinity of the
  [NeII], from data sets HD-BCK1 and HD-BCK2, obtained by taking the
  median of each of the image slices. The dotted vertical line is
  the rest wavelength of the [NeII] line.  }
 \label{fig:HDBCK12_NeII}
\end{figure}

\begin{figure}[htbp] 	
 \plotone{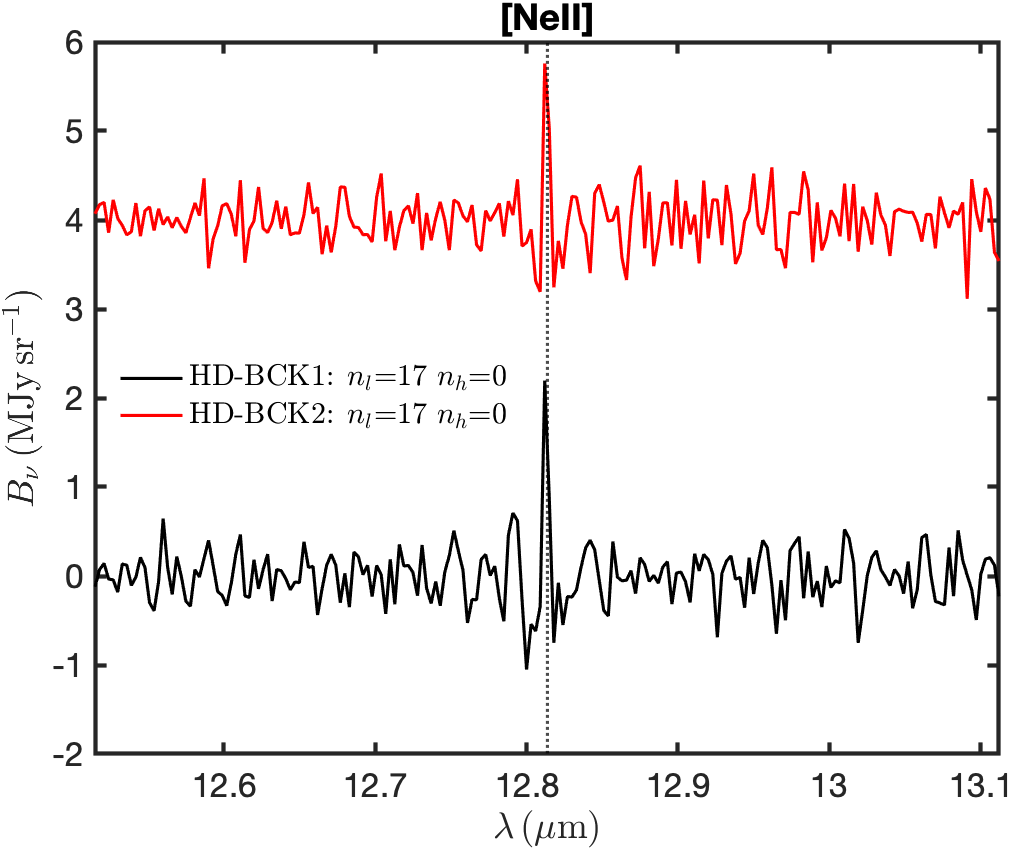} 
  \caption{\small The same as in Figure~\ref{fig:HDBCK12_NeII} but
  after Fourier filtering.   The HD-BKG2 spectrum (top) is offset from HD-BKG1 by
  $4\,{\rm MJy\,sr^{-1}}$. 
For Fourier filtering the starting point was a 
a spectrum 200-points long and centered on the line. Following Fourier transform the 
power spectrum was inspected. The first 
$n_l$ Fourier channels (low frequency) and last $n_h$ channels (high frequency)
were zeroed out and inverse transform was applied
to yield the filtered spectrum displayed here; the values
of $n_l$ and $n_h$ are noted in the figure.
  }
 \label{fig:HDBCK12_NeII_filtered}
\end{figure}

\section{Detection by MIRI-MRS}
 \label{sec:MRS_Detection}
 
Motivated by the results presented in Table~\ref{eq:SNR} we undertook
analysis of MIRI-MRS commissioning data, specifically of a calibration
star (\hd) and its associated background or ``blank" field observations
(hereafter, HD-BKG1 and HD-BKG2).  The Galactic coordinates of \hd\
is $l=89^\circ.24$ and $b=30^\circ.56$.  The blank field is some
$1.32^\prime$ North of \hd. The observing log is summarized in
Table~\ref{tab:logHD}. Summary information on \hd\ can be found in
Appendix~\ref{sec:HD163466}.  
As a cross-check of our analyses (e.g.,
in confirming inferred velocities and velocity widths) we  
also analyzed observations of the bright planetary nebula
NGC\,6543.   The observing
log and the analysis of NGC\,6543 is summarized in
Appendix~\ref{sec:NGC6543}.

For all targets, we downloaded the Level-3 MIRI-MRS pipeline data
cubes.  As a part of the pipeline reduction the  MIRI-MRS pipeline
wavelength scale has been adjusted  to  the solar system barycenter.
The data sets are cubes with images of size ($n_x,n_y$) and $n_{ch}$
equally spaced (in wavelength) spectral channels.  These numbers
vary with channel  (and dithering pattern) but are typically about
(30, 30) and 2000, respectively.

\subsection{Background datasets (HD-BKG1, HD-BKG2)}

In this sub-section we present our analysis of HD-BKG1 and HD-BGK2
data sets (see Table~\ref{tab:logHD}). For each data set the sky spectrum
is obtained by taking median of each image slice.
In Figure~\ref{fig:HDBCK12_NeII} we display the resulting spectrum
centered around [NeII] line. There seems to be a clear detection
of an unresolved line close to the rest wavelength of the [NeII]
line.  However, it is clear from the Figure,  the sky spectrum is dominated by systematics.  After inspection of this spectrum and other
spectra it became clear that the systematics are due to three
reasons: ({\it i}) periodicity in the baseline, ({\it ii}) strong
pixel-to-pixel variations and ({\it iii}) occasionally artifacts, specifically super bright resolved
lines.

The signal we are searching is an unresolved line.  
The Fourier
amplitude spectrum of an unresolved line is flat with Fourier frequency. 
This suggests a Fourier filtering scheme  which
removes the periodic
lumps in the baseline via low-frequency filtering while pixel-to-pixel
variation is removed by high frequency filtering. 

\begin{figure}[htbp]            
 \plotone{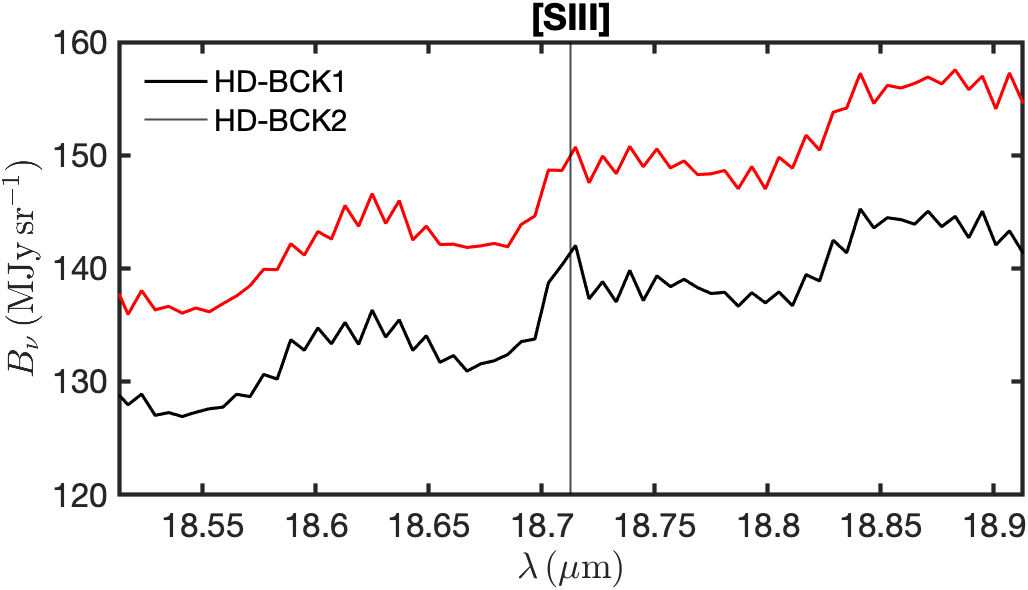}
  \caption{\small The spectra of the sky, in the vicinity of the
  [SIII], from data sets HD-BCK1 and HD-BCK2, obtained by taking the
  median of each of the image slices. The dotted vertical line is
  the rest wavelength of the [SIII] line. No vertical offsets were
  applied to the spectra. The variation is presumably due to
  incomplete calibration.  }
 \label{fig:HDBCK12_SIII}
\end{figure}


\begin{figure}[htbp] 
 \plotone{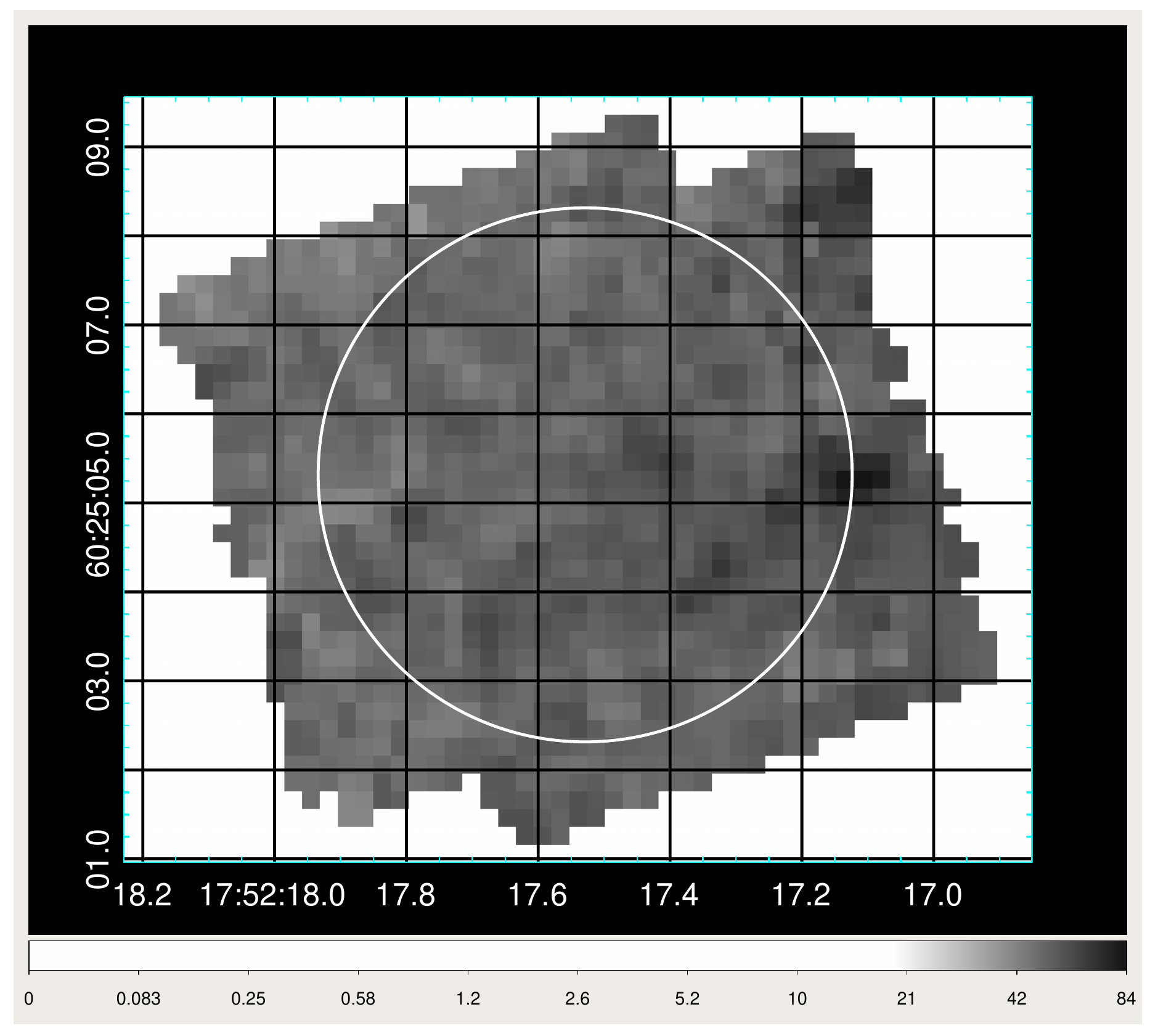}
  \caption{\small The sky image of the background region in the
  [NeII]~12.8 \mum\ line. The sky coordinates are shown in green.
  A circle with a radius of $3^{\prime\prime}$ is shown for scale.
}
 \label{fig:HD163466-BKG-Image} 
\end{figure}

As can be seen from Figures~\ref{fig:HDBCK12_NeII} and
\ref{fig:HDBCK12_NeII_filtered} the [NeII] line is convincingly
detected in the input spectrum and the filtered spectrum.
Both detections yield a similar value for the strength of the line.
We preferred to use the filtered spectrum to measure the strength
of the signal. To this end,  based on the window function used for
Fourier filtering, we synthesized a ``dirty beam", $f_D(\lambda)$.
We normalized this profile as follows: $f(\lambda)=f_D(\lambda)/\int
f_D^2(\lambda)d\lambda$. We convolved the filtered spectrum with
$f(\lambda)$. The line strength was set by the maximum amplitude.
We find the resulting line strength to be  $2.54\pm 0.31$ (HD-BCK1)
and $2.08\pm 0.30\,{\rm MJy\,sr^{-1}}$ (HD-BCK2); here, the rms was
determined from the fluctuations in the convolved spectrum (whilst
avoiding the line itself).  We adopt $2.31\pm 0.30\,{\rm MJy\,sr^{-1}}$.
Note that in Table~\ref{tab:SNR} the definition of the spectral
resolution incorporates the line width of an unresolved line,
$\mathcal{R}=\lambda/{\rm FWHM}$.  Thus, the brightness in Rayleigh
is $B_\nu/(h\mathcal{R})$ and this amounts to $15.2\pm 2.0\,R$.
The image of the sky in the [NeII] line is shown in the left panel
of Figure~\ref{fig:HD163466-BKG-Image}.

\begin{figure}[htbp]     
 \plotone{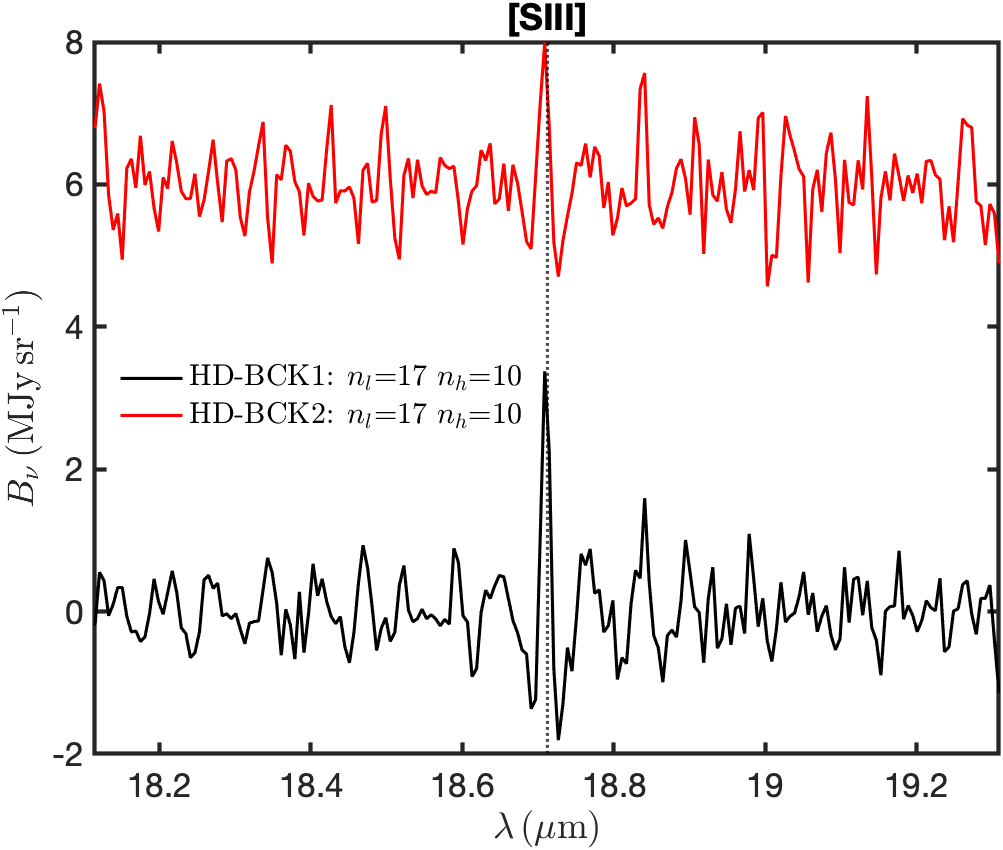} 
  \caption{\small The same as in Figure~\ref{fig:HDBCK12_SIII} but
  after Fourier filtering.  The HD-BCK2 spectrum (top) is offset from
  HD-BCK1 by $6\,{\rm MJy\,sr^{-1}}$.  }
 \label{fig:HDBCK12_SIII_filtered} 
\end{figure}

As can be seen from Figure~\ref{fig:HDBCK12_SIII}, relative to that
of [NeII], the pixel-to-variation for [SIII] is severe.
Nonetheless, by eye one can see a line at the rest velocity of
[SIII].  As can be seen from Figure~\ref{fig:HDBCK12_SIII_filtered}
the Fourier filtering nicely attenuates the pixel-to-pixel variations and flattens the baseline.
The resulting line
strengths are $3.70\pm 0.54\,{\rm MJy\,sr^{-1}}$ and $1.96\pm
0.62\,{\rm MJy\,sr^{-1}}$ which leads to a formal value of $2.83\pm
0.58\,{\rm MJy\,sr^{-1}}$.  The corresponding surface brightness
in the [SIII] line is $33.3\pm 6.8\,R$.

\begin{figure*}[htbp]           
\centering
 \includegraphics[width=0.6\textwidth]{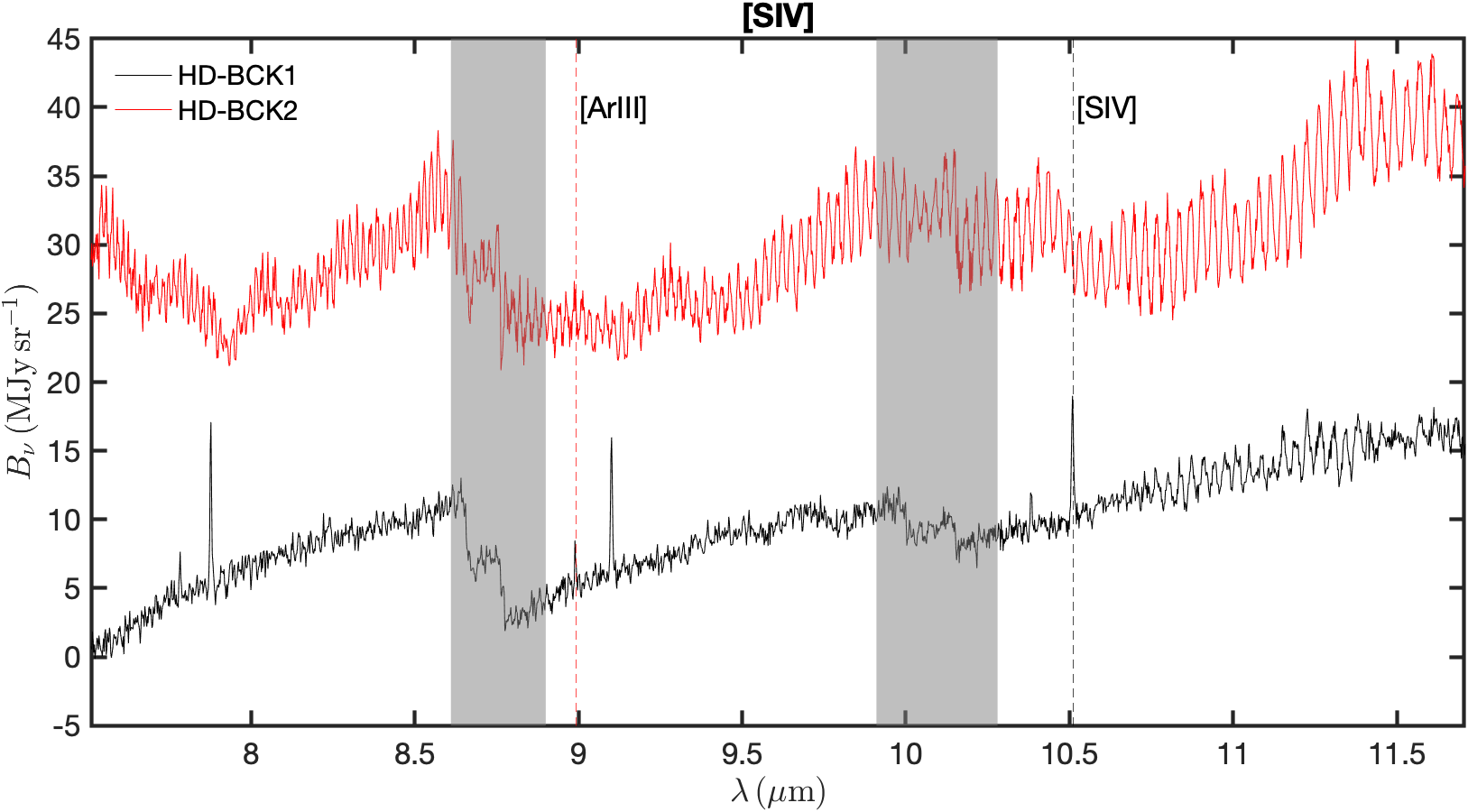}
  \caption{\small The spectra of the sky of Channel 2 (sub-bands
  A, B, C) of the data sets, HD-BKG1 and HD-BKG2.  The gray columns
  mark the overlap between adjacent sub-bands (as in A-B and B-C).
  Notice a pair of bright resolved lines in each of the sub-bands
  which is present  in HD-BKG1 (bottom) but not HD-BKG2 (top).  In
  each band, the separation is about $0.012\lambda$ with $\lambda$
  being measured in microns. The wavelength of [SIV] and [ArIII]
  are marked by vertical lines.  }
 \label{fig:HDBCK12_SIV}
\end{figure*}

\begin{figure}[hbp] 		
 \plotone{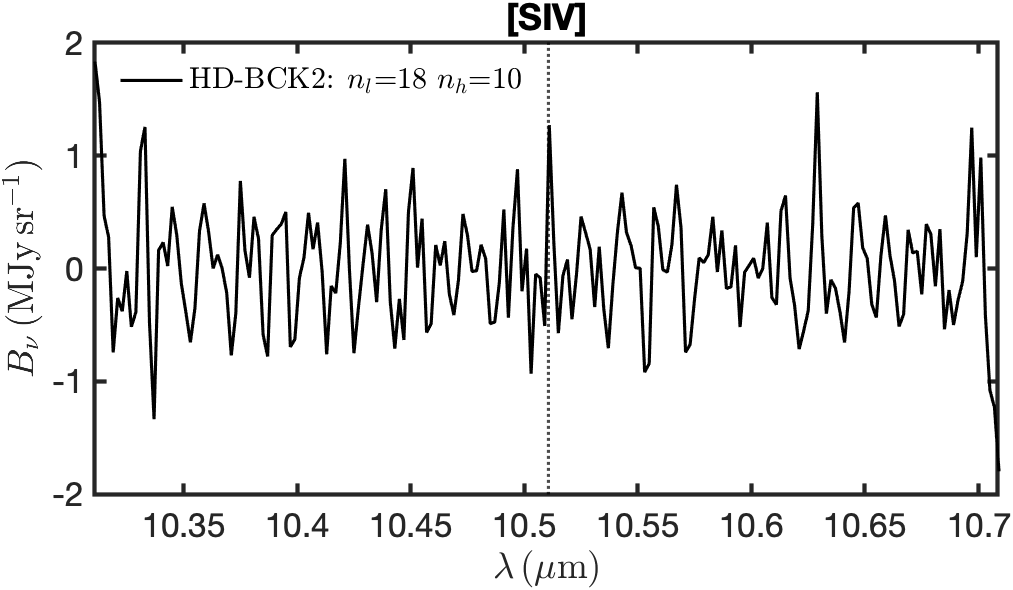} 
  \caption{\small The Fourier-filtered spectrum of the sky, in the
  vicinity of the [SIV] from HD-BCK2 data set.  The dotted vertical
  line is the rest wavelength of the [SIV] line.  }
 \label{fig:HDBCK2_SIV} 
\end{figure}

Next we consider [SIV]. As can be seen from Figure~\ref{fig:HDBCK12_SIV}
the sky spectrum of the first epoch exhibits a pair of bright lines
in each of the three bands. These lines are not present in the
second epoch. Furthermore, two of these lines coincide with [ArIII]
and [SIV]!  We suspect that these three pairs of lines are an
artifact of the calibration process. We drop data set HDBCK-1 from
further analysis.  The spectrum emerging from the analysis of HDBCK-2
is shown in Figure~\ref{fig:HDBCK2_SIV}.  The inferred 
[SIV] line intensity is $1.67\pm 0.66\,{\rm MJy\,sr^{-1}}$ which
corresponds to $10.6\pm 3.6\,R$.  No other lines were detected with
any significance. 

The summary of the detections and non-detections
can be found in Table~\ref{tab:HDBCK_results}.

\begin{deluxetable}{lrrr}[hbt]
\label{tab:HDBCK_results}
\tablewidth{0pt}
\tablecaption{Line Strengths}
\tablehead{
\colhead{line} &
\colhead{$B_\nu$ (BKG-1)} &
\colhead{$B_\nu$ (BKG-2)} &
\colhead{$I\,(R)$}
}
\startdata
{}[NeII]   & $2.54\pm 0.31$  &    $2.08 \pm 0.30$ &  $15.2\pm 2.0$\\
{}[NeIII]  & $0.63\pm 0.76$  &    $1.37 \pm 0.55$ &  $7.4\pm 4.9$\\
{}[ArII]    & $0.97 \pm 0.77$ &    $1.00 \pm 0.79$ &  $5.7\pm 4.5$\\
{}[ArIIIa] &  -                         &    $0.60 \pm 0.58$ &  $4.0\pm 3.9$\\
{}[ArIIIb] & $1.57 \pm 1.09$ &    $0.44 \pm 1.15$ &  $11.2\pm 12.5$\\
{}[SIII]    & $3.70 \pm 0.54$ &    $1.96 \pm 0.62$ &  $33.3 \pm 6.8$\\
{}[SIV]    & -                         &    $1.67 \pm 0.57$ &  $10.6\pm 3.6$
\enddata
 \tablecomments{ BKG-1 and BKG-2 refer to the two background data sets.
 The unit of $B_\nu$ is ${\rm MJy\,sr^{-1}}$.  The last column is
 the line intensity (unit: Rayleigh, $R$) obtained by taking the
 arithmetic mean of the two intensities (when available). } 
\end{deluxetable}

\subsection{\hd\ data set}

\begin{figure}[htbp]
 \plotone{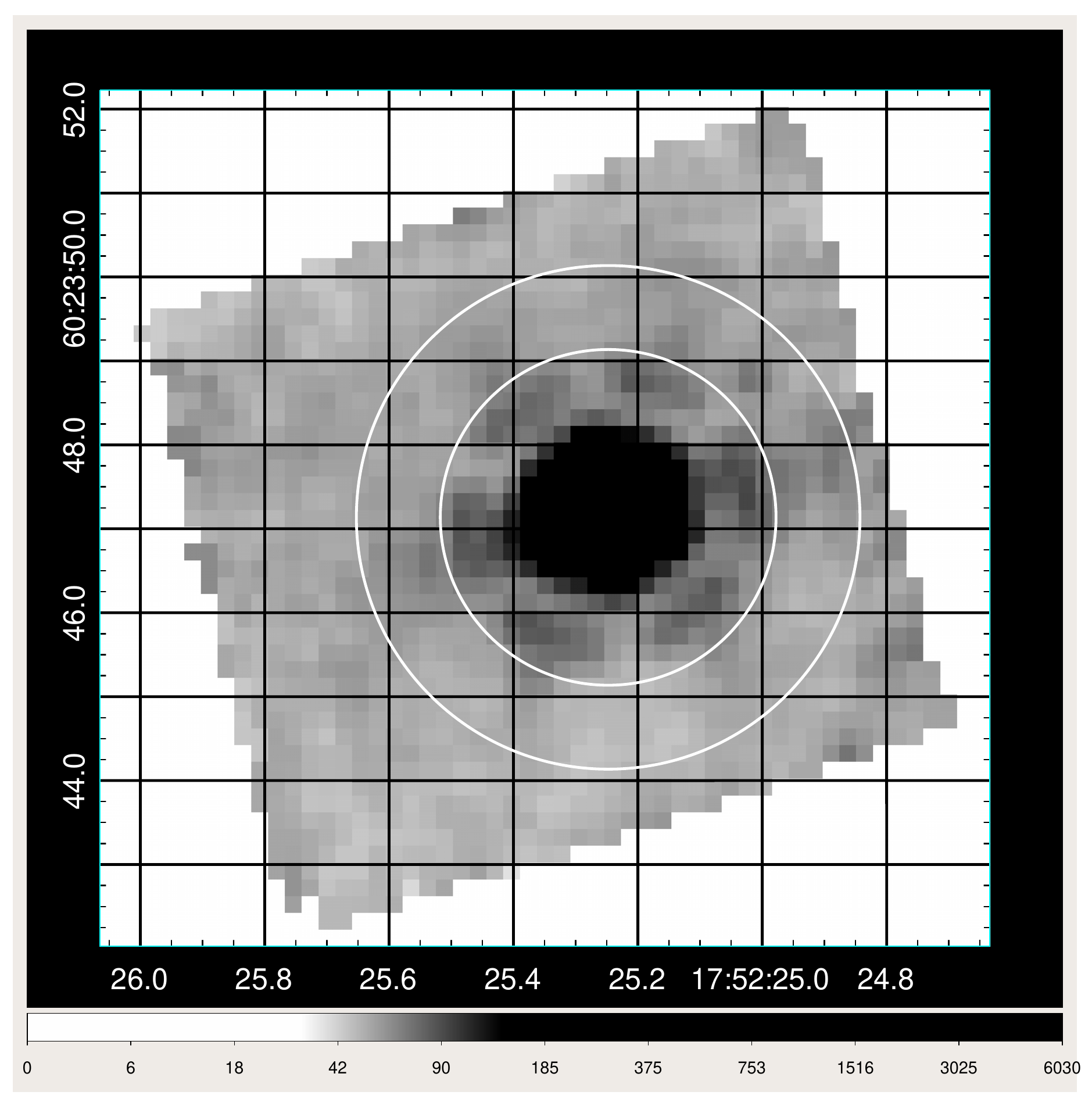}
  \caption{\small The sky image centered on \hd.  The circles define
  10 and 15-pixel radius masks used to reject light from the aperture
  around   the star \hd\ (see text).  }
 \label{fig:HD163466-Image}
\end{figure}

The data set for \hd\ is attractive because of the long exposure
time (see Table~\ref{tab:logHD}). However, this is compensated by
scattered light from \hd\ which is a bright star, $V=6.85\,$mag.
To this end, we masked out a circular region centered on the star
(see Figure~\ref{fig:HD163466-Image}) and then determined
the median of the remaining pixels. The larger the
radius of the masked region the smaller is the contribution of the light scattered from the star to
the sky spectrum but at the cost of fewer data points.
For this reason the analysis was undertaken for two
different radii (see Figure~\ref{fig:HD163466-Image}).
The results are summarized in
Table~\ref{tab:HD_dataset}.  [NeII] is robustly detected at
$\approx 1.9\,{\rm MJy\,sr^{-1}}$ and [SIII] is plausibly
detected, $S \approx 4\,{\rm MJy\,sr^{-1}}$.  The corresponding
photon intensities are $12\pm 2\,R$ and $47\pm 15\,R$. Within errors
both are consistent with the determinations from the background
fields.


\begin{deluxetable}{lrr}[bht]
\label{tab:HD_dataset}
\tablewidth{0pt}
\tablecaption{\hd\ dataset}
\tablehead{
\colhead{Line} & 
\colhead{$S (n=10)$} &
\colhead{$S (n=15)$}
}
\startdata
{}[NeII]   & $1.74 \pm 0.28$ &  $1.90 \pm 0.32$ \\
{}[NeIII]  & $1.16 \pm 0.50$ &  $1.15 \pm 0.45$ \\
{}[ArII]   & $0.82 \pm 1.22$ &  $1.32 \pm 1.54$ \\
{}[ArIIIa] & $0.45 \pm 0.61$ &  $0.80 \pm 0.77$ \\
{}[ArIIIb] & $2.58 \pm 0.97$ &  $1.75 \pm 2.64$ \\
{}[SIII]   & $0.90 \pm 0.48$ &  $4.18 \pm 1.34$ \\
{}[SIV]    & $1.49 \pm 0.73$ &  $1.01 \pm 0.82$ 
\enddata
 \tablecomments{ A circular region of radius $n$ pixels and centered
 on \hd\ of was masked out and the median of the remaining field
 used to obtain the sky spectrum. The second ($n=10$) and third
 ($n=15$) columns are the sky brightness in a single spectrometer
 channel with the unit of ${\rm MJy\,sr^{-1}}$.  }
\end{deluxetable}

\begin{figure}[htbp] 
 \plotone{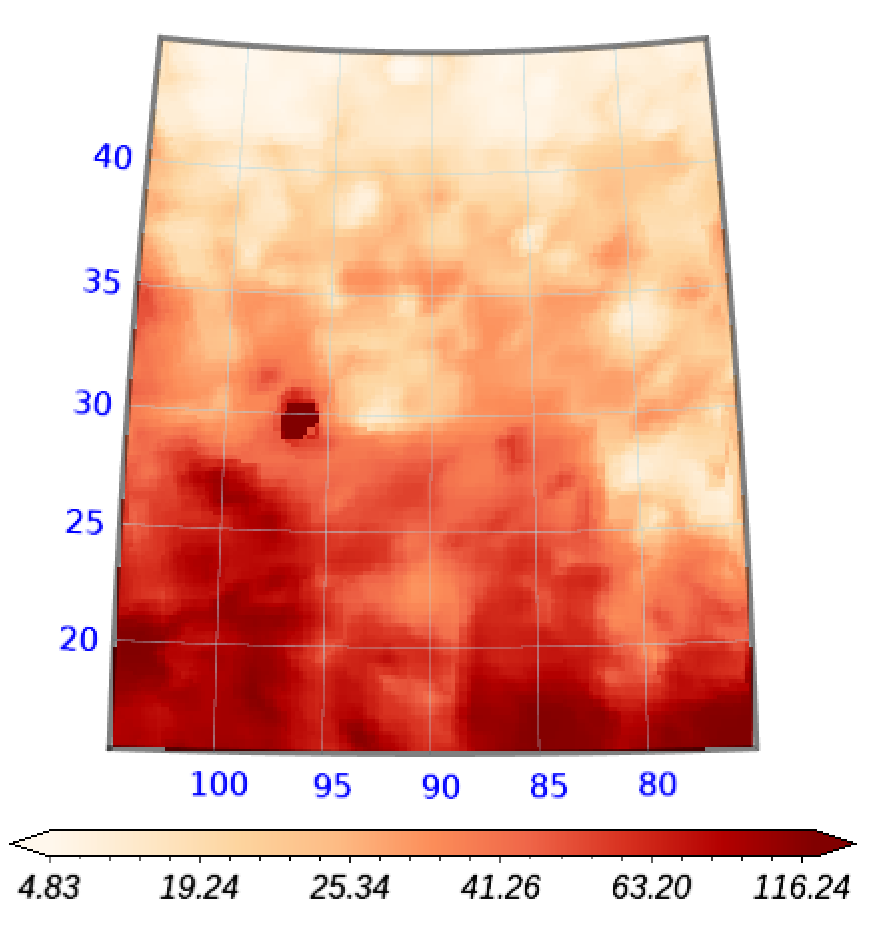}
  \caption{\small WHAM H$\alpha$ wide image, in Galactic coordinates,
  of the sky centered on \hd\ (located at $l=89^\circ.24$,
  $b=30^\circ.56$). The horizontal ``thermometer" is EM (${\rm cm^{-3}\,pc}$) scaled up by a factor of 22. In the direction towards
  HD\,163466 (marked by a circle) the
  emission measure is about 
  $1\,{\rm cm^{-6}\,pc}$.
  The bright emission region
  at $l\approx 95^\circ$ and $b\approx 30^\circ$ is NGC\,6543.
  Figure provided by M.\ Haffner.}
 \label{fig:WHAM_widefield}
\end{figure}

\section{Inference}
 \label{sec:Inference}

The [NeII]\,12.814\,$\mu$m photon intensity of $15.2\pm 2\,R$ (see
Table~\ref{tab:HDBCK_results}) can, using Equation~\ref{eq:NeII},
be used to infer the EM, subject only to our assumption of temperature,
$T$ of 8,000\,K and adopted gas-phase abundance of neon.  We make
little error in assuming $x_{\rm H^+}=1$. With that simplification
the emission measure is $8.2x_{\rm Ne^+}^{-1}$. Even if $x_{\rm
Ne^+}$ is as low as 0.5 the inferred emission measure is modest,
${\rm 16\,cm^{-6}\,pc}$.  Thus, it is  reasonable to assume that
[NeII] emission arises in the WIM. In support of this inference we note
that the [NeII] emission towards \hd, some 1.3$^\prime$ away, is
at the same level as that measured in the background field direction.
From Table~\ref{tab:HDBCK_results} we find that the 2-$\sigma$ upper
limit is [NeIII]/[NeII]$<1.13$ and is not particularly informative.

\begin{figure}
 \plotone{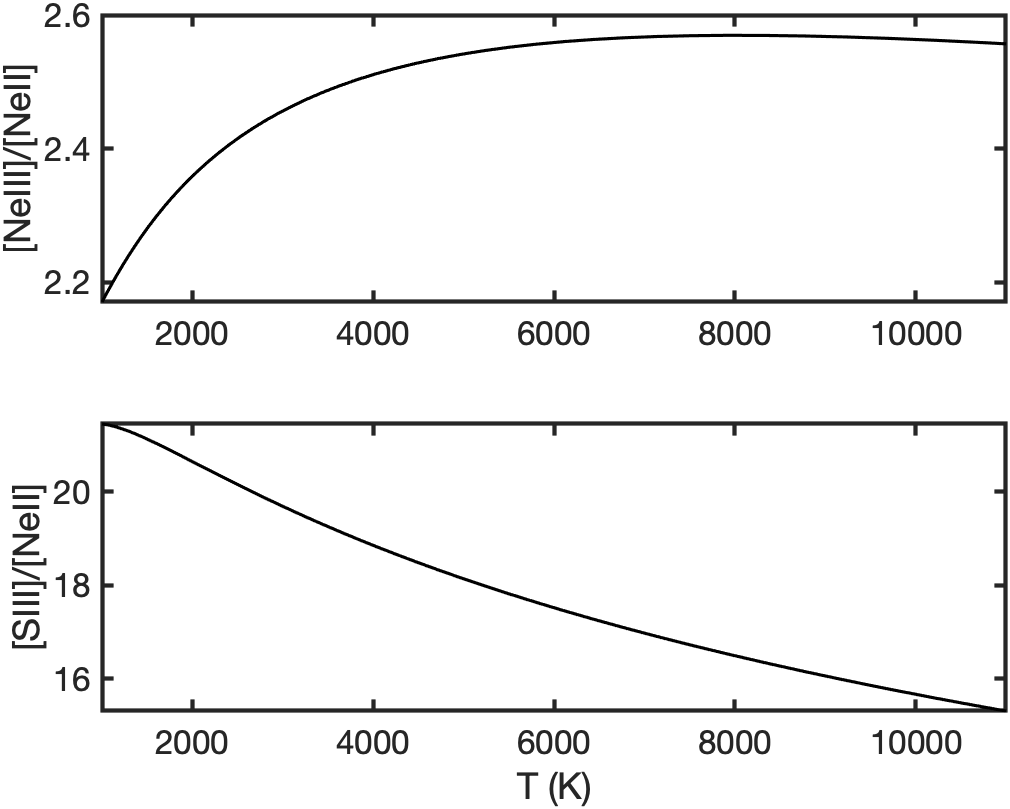}
    \caption{\small 
    (Top) The run of [NeIII]/[NeII] as 
    a function of temperature.
    (Bottom) The run of [SIII]/[NeII] as a 
    function of temperature. The multiplicative scale factor is 
    $x_{\rm S^{+2}}/x_{\rm Ne^+}$.
    }
    \label{fig:Neon_ratio}
\end{figure}

One of the four instruments on the 
Infrared Space Observatory (ISO) was the
Short Wave Spectrometer (SWS; 2.4--45\,$\mu$m).
\citet{gsl+02} undertook an extensive
study of HII regions in a number of mid-IR fine structure lines. They found  [NeIII]/[NeII] to vary from 0.1 (at small Galactocentric radii) to unity
(at the solar circle and beyond), albeit with significant scatter.
For \hd\ the observed [NeIII]/[NeII] ratio is $0.49\pm 0.32$. Given that we have a single
determination and the scatter in
the radial dependence of the [NeIII]/[NeII] ratio we are not able to draw any useful
conclusion. Next, 
as can be seen from Figure~\ref{fig:Neon_ratio}
the [NeIII]/[NeII] intensity ratio is quite insensitive to
the temperature of the gas. At the 3-$\sigma$ level we see that, $x_{\rm
Ne^{+2}}/x_{\rm Ne^+}<0.57$ which means that [NeII] is more abundant
than [NeIII] -- this is entirely expected in ionization models of
the WIM (e.g., \citealt{M86}).

In Figure~\ref{fig:Neon_ratio} we plot
the [SIII]18.71/[NeII]12.81 ratio. The observed
ratio is $2.2\pm 0.44$ from which we deduce that
$x_{\rm S^{+2}}/x_{\rm Ne^+}=0.133\pm 0.027$.
To the extent that our adopted
abundances are accurate this result points to a significant decrease
in the ionizing power of the incident EUV field from 21.6 eV
(ionization potential of Ne~I; see Table~\ref{tab:IP_abundance})
to 23.3 (the ionization potential of S~II;
{\it ibid}).

\subsection{H$\alpha$: WHAM}
 \label{sec:Halpha}
 
WHAM produced an H$\alpha$ map of the sky at one-degree angular
resolution \citep{hrt+03}.  The WHAM H$\alpha$ map centered on \hd\
is shown in Figure~\ref{fig:WHAM_widefield}.  As can be seen from
Figure~\ref{fig:WHAM_widefield} the field of \hd\ is not dominated
by any strong nebula.

\begin{figure}         
 \plotone{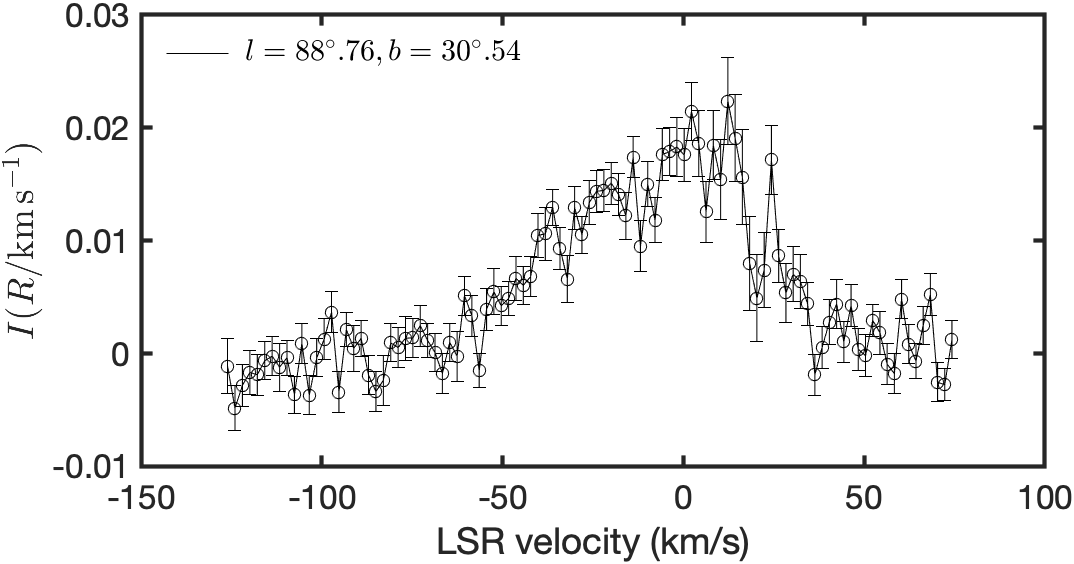}
  \caption{\smallskip The H$\alpha$ intensity spectrum from the
  Wisconsin H$\alpha$ Mapper (WHAM). The unit for intensity  is
  $R$\,(km\,s$^{-1})^{-1}$ while the $x$-axis is the velocity with
  respect to the local standard of rest (LSR).  The spectrum is
  typical in this region and is usually decomposed into into two
  Gaussian components: one centered at $v_{\rm LSR}\approx 0\,{\rm km\,s^{-1}}$
  for the local (Orion) arm and the other at $\approx -30\,{\rm
  km\,s^{-1}}$ (Perseus arm).}
 \label{fig:WHAM_spectrum}
\end{figure}

From the WHAM catalog\footnote{\url{http://www.astro.wisc.edu/wham-site/};
see \citep{hrt+03}} we extracted the spectrum closest to the
line-of-sight to \hd\ (see Figure~\ref{fig:WHAM_spectrum}).  The
integrated intensity is  $1.14\pm 0.04\,R$.  The peak at $v_{\rm
LSR}\approx 0$ arises in the local ``Orion" spur while the negative
velocity emission at, say, $v_{\rm LSR} \approx {\rm -30\,km\,s^{-1}}$
arises in the Perseus arm (see \citealt{xrd+16}).  The intensity
weighted mean LSR velocity is $-2\,{\rm km\,s^{-1}}$.  The
estimated\footnote{\url{https://irsa.ipac.caltech.edu/applications/DUST/}}
reddening, ${\rm E(B-V)}$, is about 0.05\,mag. The corresponding
extinction at H$\alpha$ is about 0.1\,mag.  The extinction corrected
value is $1.26\pm 0.04\,R$.  Some of the observed H$\alpha$ is due
to scattering of brighter Galactic H$\alpha$ by interstellar dust
into the line-of-sight.  We apply a scattering correction of 15\%
\citep{wr99} and find $I_{\rm H\alpha}=1.07\pm 0.03\,R$.  For
$T=8,000\,$K, the inferred emission measure is $2.4\,{\rm cm^{-6}\,pc}$
significantly smaller than that inferred from observations of [NeII]. Given the widely differing angular
scales\footnote{nearly a factor of $10^6$ in solid angle} of MRS
and WHAM (cf.\ \S\ref{sec:IonizationFraction}), we are not alarmed
by the discrepancy. 

\subsection{[NeII]}

The barycentric velocity of the [NeII] line is $-20\pm 12\,{\rm
km\,s^{-1}}$ (also Table~\ref{tab:datafits}).  The corresponding
velocity with respect to the Local Standard of Rest (LSR) is
$-37\pm12\,{\rm km\,s^{-1}}$.  It appears that the [NeII] emission
is associated with negative velocity H$\alpha$ emission which can
be traced to the  Perseus arm, at a distance of about 5\,kpc (see
\citealt{xrd+16}).  The vertical scale height is 2.5\,kpc. The
temperature of the WIM rises with the vertical distance
\citep{hrt99,mrh06}. It may well be that a better value for the
temperature is $10^4\,$K -- which, for a given EM, would increase
[NeII] and [SIII] emission.

H$\alpha$ measurements obtained on angular scales similar to that
of MRS, if available, would have allowed us to infer the ionization
fraction of Ne$^+$ and S$^{++}$. We currently lack this option
because the primary source of H$\alpha$ data is WHAM which has
one-degree beam. Fortunately, as discussed below (\S\ref{sec:JointStudies})
modern IFUs on large ground-based optical telescopes will make it
possible to measure not only H$\alpha$ but also the full suite of
optical nebular lines of interest to WIM studies.

\section{Conclusions}
 \label{sec:Conclusions}

MIRI-MRS, as a part of commissioning  observed a calibrator star,
\hd\ and an associated ``background" field. This intermediate latitude line-of-sight ($l=88^\circ.8$, $b=30^\circ.5$) does not
include any bright nebulae.  We obtained the sky spectrum by taking
the median of each image slice of the background field data cube.
We detected strong [Ne]\,$12.814\,\mu$m at a barycentric velocity
of $-20$\,km\,s$^{-1}$,  [SIII]\,18.713\,$\mu$m and possibly
[SIV]\,10.510\,$\mu$m. The photon statistics in the \hd\ data set
are dominated by photons emitted by bright calibrator star.  We
masked out bright emission from the stars and undertook a similar
analysis. [NeII] emission is readily detected while [SIII] is
marginally detected.  From the neon observations we infer an emission
measure, EM$\approx (8.2\pm 1.1)x_{\rm H^+}/x_{\rm Ne^+}\,{\rm
cm^{-3}\,pc}$.  The low value of the inferred EM lead us to conclude
that these lines arise in the Galactic WIM.  We measure $x_{\rm
Ne^+}/x_{\rm Ne^{++}}<0.6$ and $x_{\rm S^{++}}/x_{\rm Ne^+}=0.13\pm
0.027$. These values are consistent with the standard low
ionization parameter model for the WIM \citep{M86,dm94,shr+00} On
a one-degree scale (as compared to few arcsecond scale of MRS) the
inferred EM from ground-based H$\alpha$ emission is  $2.4\,{\rm
cm^{-3}\,pc}$.

The detection of [NeII] emission on arcsecond scales augers well
for the MRS to study tiny ionized nebulae.  The radius of the
\Stromgren\ sphere is
 \begin{equation*}
  R_S =\bigg(\frac{3Q_0}{4\pi n_H^2\alpha}\bigg)^{1/3} = 1.5\times
  10^{-2}\,Q_{40}^{1/3}n_1^{-2/3}\,{\rm pc}
 \end{equation*}
where  $n_{\rm H}=10n_1\,{\rm cm^{-3}}$ is the H-atom density, $Q_0$
is the rate of ionizing photons emitted by the star and
$Q_{40}=Q_0/10^{40}\,{\rm photon\,s^{-1}}$.  The corresponding emission
measure is $ {\rm EM} =1.94n_1^{4/3}Q_{40}^{1/3}\,{\rm cm^{-6}\,pc}$.
The ionizing photon luminosity of a white dwarf (radius,
$R=0.014\,R_\odot$) at a temperature $T=[1.7\times 10^4, 1.0\times
10^4]\,$K is  $[2.9\times 10^{40}, 2.8\times 10^{37}]\,{\rm s^{-1}}$.
So a white dwarf embedded in the Warm Neutral Medium (volume filling
factor of over a third; $n_{\rm H}\approx 1\,{\rm cm^{-3}}$) or an
A star embedded in the Cold Neutral Medium (filling factor less
than a percent; $n_{\rm H}\approx 30\,{\rm cm^{-3}}$) will have
nebulae that are detectable by MRS.  Both nebulae lack sharp
\Stromgren\ spheres.

\subsection{A JWST Comensal Program}
 \label{sec:Comensal}
 
The results presented here are limited by uncorrected fringing and not fully calibrated
spectral baselines. The current state of affairs is not surprising.
IFUs are powerful but are also complex instruments. Past experience
has shown that it will take effort and time to calibrate IFUs.  We
will assume that in due course the MIRI-MRS pipeline will routinely
produce photon-noise limited data cubes and the sensitivities
summarized in Table~\ref{tab:SNR} will be achieved.

The median WIM H$\alpha$ emission for $\vert b\vert>20^\circ$ is
$1\,R$ which corresponds to an emission measure of $2.2\,{\rm
cm^{-3}\,pc}$.  Several large regions of low-latitude sky (e.g.,
Perseus arm, $l=130^\circ, b=-7.5^\circ$; see \citealt{mrh06}) have
bright WIM emission, $I_{\rm H\alpha}\approx 10\,R$.  Table~\ref{tab:SNR}
shows that with an hour long integration MIRI-MRS has the capability
to detect [NeII], [ArII], [SIII] and perhaps even [SIV] from a good
fraction of the sky, both at high latitude and most certainly at
intermediate and low latitude.

Thus, over the course of the lifetime of JWST, we can expect suitable
MRS data sets to grow steadily. This is a fine example of highly
productive ``comensal" observing by JWST.  The increasing sample
will give astronomers statistical power to explore the variations in the
spectrum of the diffuse EUV radiation field and, in due course, 
measure the volume filling factor of the Warm Neutral Medium (WNM).

\subsection{Joint studies with ground-based IFUs}
 \label{sec:JointStudies}

Ground-based optical facilities are well
suited to studies via nebular lines as well as hydrogen and helium
recombination line studies.  Over the last decade, high spectral
resolution IFU spectrographs have been commissioned on large
ground-based optical telescopes (e.g, MUSE/VLT, \citealt{bvb+14};
KCWI/Keck, \citealt{mmm+18}).

The sky background, on a moonless and clear night, at Paranal
\citep{jnk+13} or Mauna
Kea\footnote{\url{https://www.gemini.edu/observing/telescopes-and-sites/sites\#OptSky}}
is about $200\,{\rm photon\,m^{-2}\,\mu{m}^{-1}\,arcsec^{-2}}$ for
for $\lambda<0.7\mu$m which corresponds to about $1.1\,R$\,\AA$^{-1}$.
The background gradually rises to ten times this value as one
proceeds to 1\,$\mu$m. For some lines, air glow line emission will
dominate over the continuum background.  Geo-coronal H$\alpha$
emission is 2--4\,$R$ even at mid-night \citep{nmr+08}. [NI]  and
[OI] lines are both very strong (up to 30\,$R$ for [NI] and up to
200\,$R$ for [OI]) and variable.  The air glow lines will be at zero
topocentric velocities while the velocity of the astronomical signal
will include the projected component of earth's orbital velocity.
Thus, for some lines of sight, with some planning, the air glow
contribution can be reduced.

Let $\mathcal{B}$ be the sky brightness in unit of Rayleigh per
Angstrom and $\mathcal{S}$ be the photon intensity of the line of
interest, also in Rayleigh.  We adopt 1\,\AA\ as the default
spectral channel width, $\Delta\lambda$ and separately note that
at the wavelength of  H$\alpha$, for instance, 1\,\AA\ corresponds
to 45\,km\,s$^{-1}$.  The signal-to-noise ratio (SNR) for an IFU
with field-of-view of $\Omega$ operating in a light bucket mode is
  \begin{equation}
	{\rm SNR} = \frac{\eta\alpha\mathcal{S}A\Omega t}
	{(\eta\alpha\mathcal{B}A\Omega t\Delta\lambda)^{1/2}}
  \end{equation}
where $\alpha=10^6/(4\pi)$ is the conversion factor between intensity
in Rayleigh to photon intensity and the other symbols have the same
meaning as in Equation~\ref{eq:SNR}.

An illustrative example is the Keck Cosmic Imager.\footnote{This is
a two-armed spectrograph; the working blue arm 
\citealt{mmm+18})  and the soon-to-be-commissioned red arms.} At the highest spectral resolution, the input
aperture is $\Omega =20^{\prime\prime}\times 8.2^{\prime\prime}$.
We simplify by setting $\eta=0.15$ and $\mathcal{R}=15,000$ for the
entire optical range. The spectral FWHM would then vary from 0.2\,\AA\
at the blue end to 0.6\,\AA\ at the red end.  The signal-to-noise
ratio (SNR) is
 \begin{equation} {\rm SNR}=361 \bigg(
  \frac{\mathcal{S}}{\sqrt{\mathcal{B}}}\bigg)
  \Big(\frac{\eta}{0.15}\Big)^{1/2}
  \Big(\frac{\Delta\lambda}{1\,\textrm{\AA}}\Big)^{-1/2} t_{\rm
  hr}^{1/2}{\rm EM}
 \end{equation}
where $t_{\rm hr}$ is the integration time in hours.  H$\alpha$
data, when combined with JWST [NeII] detection, will make it possible
to deduce ionization fraction of Ne$^+$. Furthermore, this SNR is
so high that useful images can be constructed with 1-arcsecond
pixels.  Thus, ionized nebulae can be probed on arc-second scales!

\begin{figure}[htbp] 
 \plotone{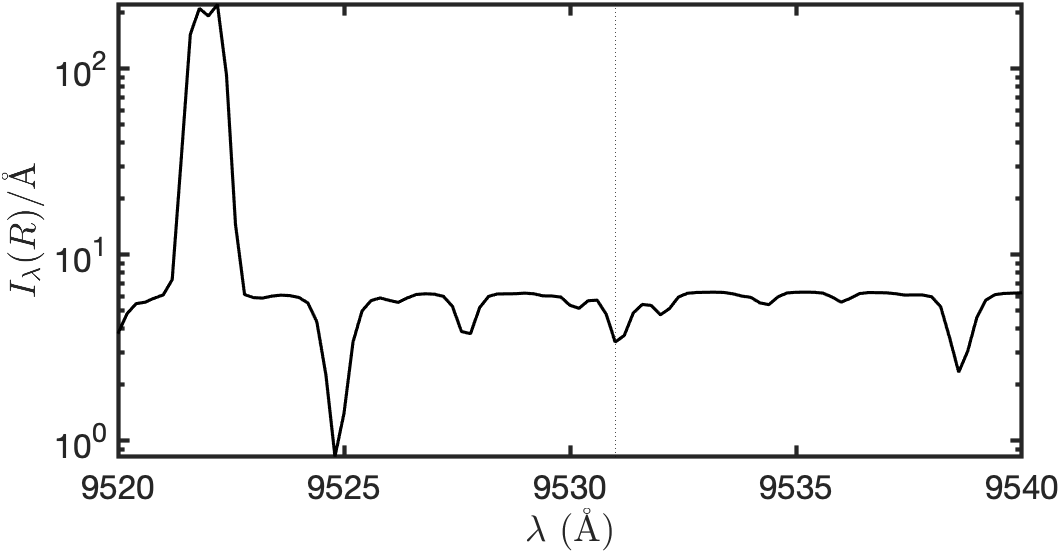}
  \caption{\small The spectrum of the sky in the vicinity of the
  [SIII]\,9531\,\AA\ line (marked vertical dotted line), as observed at the Gemini Observatory
  atop Mauna Kea.  }
 \label{fig:SIII_NIR} 
\end{figure}

Next, consider sulfur and neon. The optical nebular lines are temperature
sensitive.  The sky
spectrum\footnote{\url{http://www.gemini.edu/sciops/ObsProcess/obsConstraints/atm-models/skybg_50_10.dat}}
in the vicinity of [SIII]\,$\lambda$9531\,\AA\ line is shown in
Figure~\ref{fig:SIII_NIR}.  The expected SNR for this line $304\xi_{\rm
S^+2}/x_{\rm H^+}t_{\rm hr}^{1/2}{\rm EM}$.  The SNR for
[NeIII]\,$\lambda$3869\,\AA\ is $90\xi_{\rm Ne^+2}/x_{\rm H^+}
t_{\rm hr}^{1/2}{\rm
EM}$.  Joint analysis of the optical and MIRI-MRS data will yield
the temperature of the WIM.  In addition, the traditional optical
forbidden lines (e.g., [OII], [OIII], [NII], [SII] etc) can be
observed with KCI and thereby obtain a comprehensive view of the
ionization of the WIM.  Inversely, the large SNRs for many optical
lines mean that every deep KCI observation is comensal -- with every
deep high spectral resolution observation leading to detections of
the WIM.

%

\subsection{Helium Recombination Lines}
 \label{sec:HeliumRecombination}

Pinning down the ionization fraction of helium in the WIM is a long
standing goal. The two strongest recombination lines
are 5865\,\AA\ line and 1.083\,$\mu$m. \citet{bss99}
provide volume emissivity for various helium recombination lines.
These emissivities are referenced to the emissivity
of the He~I\,$\lambda$\,4471\,\AA\ ($1s4d\,^3{\rm D}\rightarrow
1s2p\,^3{\rm P}^o$ line.\footnote{In the triplet system this is the equivalent
of H$\beta$.} We find the volume emissivities for
HeI\,$\lambda$1.083\,$\mu$m triplet ($1s2p\,^3{\rm P}^o\rightarrow
1s2s\,^3{\rm S}$) and HeI\,$\lambda$5875\,\AA\ ($1s3d\,^3{\rm
D}\rightarrow 1s2s\,^3{\rm S}$) to be the following:
 \begin{eqnarray*}
   4\pi n_{5875} &=& 5.0\times 10^{-14}T_4^{-1.065}n_en_{\rm
   H^+}\,{\rm cm^3\,s^{-1}}\\ 4\pi n_{1.083} &=& 1.9\times 10^{-13}
   T_4^{-0.523} n_e n_{\rm H^+}\,{\rm cm^3\,s^{-1}}
 \end{eqnarray*}
where $\dot{n}_\lambda=j_\nu/(h\nu)$ with $j_\nu$ being the emissivity.
The corresponding  photon intensity along a line-of-sight is
 \begin{eqnarray*}
    I_{5875} &=& 1.26\times 10^{-2}\xi_{\rm He^+}T_4^{-1.065}\,{\rm
    EM}\,R \ ,\\
   I_{1.083}& = & 4.8\times 10^{-2}\xi_{\rm He^+}T_4^{-0.524}\,{\rm
   EM}\,R\  .
  \end{eqnarray*}
A few observations were undertaken with WHAM in HeI~$\lambda$5875
\citep{rt95}. Here, we explore the detectability of the latter with
NIRSpec/JWST operating in the IFU mode \citep{bal+22,jfo+22}.

For the IFU we assume an entrance aperture, $\Omega= 3^{\prime\prime}\times
3^{\prime\prime}$.  For HeI~1.083\,$\mu$m line the appropriate
grating is G140H, $\eta\approx 0.3$ and $\mathcal{R}=2700$
\citep{gbb+22}. The corresponding background intensity is ${\rm
0.3\,MJy\,ster^{-1}}$.  For $T=8,000\,$K the HeI\,1.083\,$\mu$m
intensity is $0.054\xi_{\rm He^+}\,{\rm EM}\,R$.  The resulting SNR
is
 \begin{equation*}
	{\rm SNR} = 2.5\xi_{\rm He^+}\Big(\frac{\eta}{0.3}\Big)^{1/2}
	\Big(\frac{\mathcal{R}}{2700}\Big)^{1/2}\Big(\frac{t}{1\,{\rm
	hr}}\Big)^{1/2} \,{\rm EM}\ .
 \end{equation*}
Directions with EM of $\gtrsim 10\,{\rm cm^{-6}\,pc}$ are ideal to
measure the helium ionization fraction.  Over the duration of the
mission, stacking of pointings with EM of few ${\rm cm^{-6}\,pc}$
would probe the ionization at high latitude.

\begin{figure}[htbp]            
 \plotone{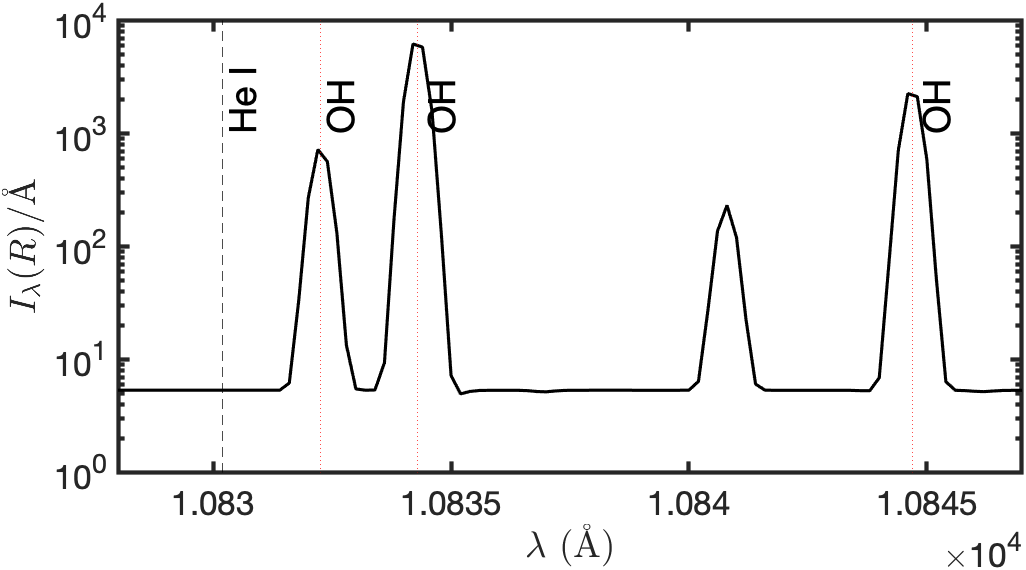}
  \caption{\small The spectrum of the sky as observed at the Gemini
  Observatory atop Mauna Kea.  The wavelengths are measured in air.
  The black vertical dotted line marks the position of HeI\,1.08300\,$\mu$m
  line.  The  red dotted lines mark OH lines. These are, from left
  to right, OH 5-2 Q2 (0.5), 5-2 Q1(1.5), 5-2 Q1(2.5); see
  \citet{rlc+00}.  }
 \label{fig:HeI_OH}
\end{figure}


How about ground-based facilities? In the vicinity of the 1.083\,$\mu$m
recombination line the sky continuum intensity is about $5\,R$\,\AA$^{-1}$
(see Figure~\ref{fig:HeI_OH}). He atoms in the terrestrial exosphere
are excited by energetic electrons to the $1s2p\,^3{\rm P}$ meta stable
state and these resonantly scatter solar HeI\,1.083\,$\mu$m photons.
The emission decreases as the sun sets.  A key requirement is a
spectral resolution of better than 1\,\AA\ (to avoid contamination
by super-strong OH lines; see Figure~\ref{fig:HeI_OH}). A spectral
resolution of 0.5\,\AA\ is even better (to eliminate terrestrial
He~I emission); thus, $\mathcal{R}\gtrsim 2\times 10^4$.   Consider
the following hypothetical (future) instrument on an 8-m telescope:
$\Omega=100\,$arcsec$^2$, $\eta=0.3$, $\mathcal{R}=2\times 10^4$.
The SNR is then $9.7\xi_{\rm He}t_{\rm hr}^{1/2}{\rm EM}$. Thus, such an
instrument would be more sensitive than NIRSPEC/JWST. Another
possibility is a Fabry-P\'erot imager (on a smaller telescope).
A different approach is to use optical lines, 
specifically the
$\lambda$5865 line. With KCWI the SNR is 
$10.4\xi_{\rm He}t_{\rm hr}^{1/2}$.

\subsection{Fine Structure in the WIM}

At the beginning of this section we noted that the inferred EM from [NeII] is
$(8.2\pm1.1)\xi_{\rm Ne^+}^{-1}$. In \S\ref{sec:Halpha} we concluded that 
the emission measure inferred from the 1-degree
WHAM H$\alpha$ sky survey, after correcting for reflection from
dust, is ${\rm EM}=2.4\,{\rm cm^{-6}\,pc}$. 
It appears that
the EM at the smaller angular scale of MIRI-MRS (a few arcseconds)
is $3.4/\xi_{\rm Ne^+}$ larger than that measured on one degree
scale. 
We advance the idea that the WIM has significant
fluctuations on sub-degree scales which are smoothed over by
the degree-beam of WHAM.  If so, in due course, comparison of
MIRI-MRS observations with WHAM will provide insight into the fine structure of the WIM. 

\acknowledgements

We are grateful to Bruce Draine, Princeton University and Ron
Reynolds, University of Wisconsin for providing discussions and
feedback and Matt Haffner, Embry-Riddle Aeronautical University,
for discussions of the WHAM data. We thank
Less Armus, IPAC, for discussions regarding 
spectral baselines.
SRK thanks Bryson Cale, IPAC-Caltech,
for help with Python coding.  Some of the research described in
this publication was carried out in part at the Jet Propulsion
Laboratory, California Institute of Technology, under a contract
with the National Aeronautics and Space Administration (80NM0018D0004).

 \bibliography{neon}{}
\bibliographystyle{aasjournal}

\appendix

\section{Fine-structure lines}
 \label{sec:AtomicData}
 
\begin{deluxetable}{lllllll}[htb]
 \tablecaption{Fine-structure lines}
\label{tab:Atomic_FSL}
\tablewidth{0pt}
\tablehead{
 \colhead{ion} &
 \colhead{$u$--$l$} & 
 \colhead{$\lambda\ (\mu{\rm m})$}  &
  \colhead{$A_{ul}\,({\rm s}^{-1})$} &
  \colhead{0} &
 \colhead{$T_{u0}$\,(K)} & 
 \colhead{$\Omega_{u0}$}
 }
\startdata
{{[}NeII]} & $^2{\rm P}_{\sfrac{1}{2}}^o$--$^2{\rm P}_{\sfrac{3}{2}}^o$ 
& 12.813548(20) &  $8.6\times 10^{-3}$ 
& $^2{\rm P}_{\sfrac{3}{2}}^o$ & 1123
& $0.314T_4^{0.076+0.002{\rm ln}\,T_4}$\\
\hline
{{[}NeIII]} & $^3{\rm P}_1$--$^3{\rm P}_{2}$ 
& 15.5551     & $5.8\times 10^{-3}$ 
& $^3{\rm P}_2$  & 925 
& $0.774T_4^{0.068-0.0556\,{\rm ln}\,T_4}$\\ 
{{[}NeIII]} & $^3{\rm P}_0$--$^3{\rm P}_{1}$ 
& 36.0135$^a$ & $1.1\times 10^{-3}$
& $^3{\rm P}_2$ & 1324      
& $0.208T_4^{0.056-0.053\,{\rm ln}\,T_4}$\\
\hline
{{[}ArII]} & $^2{\rm P}_{\sfrac{1}{2}}^o$--$^2{\rm P}_{\sfrac{3}{2}}^o$ 
& 6.985274(4)   & $5.3\times 10^{-2}$ 
& $^2{\rm P}_{\sfrac{1}{2}}^o$  & 2060
& $2.93T_4^{0.084-0.014{\rm ln}\,T_4}$\\
\hline
{{[}ArIII]} & $^3{\rm P}_1$--$^3{\rm P}_2$  
& 8.99138(12) & $3.1\times 10^{-2}$ 
& $^3{\rm P}_2$  & 1601
& $4.04T_4^{0.031+0.002{\rm ln}\,T4}$       \\
{{[}ArIII]} & $^3{\rm P}_0$--$^3{\rm P}_1$
& 21.8302(3)& $5.9\times 10^{-3}$ 
& $^3{\rm P}_2$ & 2259 
& $1.00T_4^{0.111-0.009{\rm ln}T_4}$           \\
\hline
{{[}SIII}] & $^3{\rm P}_2$--$^3{\rm P}_1$ 
&  18.713   &  $2.6\times 10^{-3}$ 
& $^3{\rm P}_0$ & 1199
&  $7.87T_4^{-0.171-0.033{\rm ln}\,T_4}$\\
\hline
{{[}SIV]} & $^2{\rm P}_{\sfrac{3}{2}}^o$--$^2{\rm P}_{\sfrac{1}{2}}^o$  &
10.5105 & $7.4\times 10^{-3}$  
& $^2{\rm P}_{\sfrac{1}{2}}^o$   & 1369
& $8.54T_4^{-0.012-0.076{\rm ln}\,T_4}$
\enddata
 \tablecomments{\small  The  quoted vacuum wavelength is 
 the one with uncertainty of the Ritz
 or measured values. The uncertainty,
 shown in the parenthesis, is the last one or two digits of the
 wavelength value.  Column 1 is the species and column 2 is
 the terms of the transitions (with $l$ for lower state and $u$ for the upper state).
 The wavelengths (column 3) and A-coefficients (column 4) are from NIST.
 The column marked with ``0" is the term for the ground level.  $T_{u0}=E_{u0}/k_B$ is the line energy equivalent temperature
 (column 6); here,
 $E_{u0}$ is the energy of the upper level $u$ (noted in the
 first column) with respect to ground level (``$0$'').  The collisional
 strength (last column), $\Omega_{u0}$, is for excitation from level $0$ to level
 $u$; the  data are from \cite{D11}.  $^a$This line is beyond the reach of
 the spectrometers of JWST but is include here because excitations
 to $^3{\rm P}_0$ also result in emission of 15.56\,$\mu$m photons.
 }
\end{deluxetable}

\section{\hd}
 \label{sec:HD163466}

\noindent \hd\ ($\alpha$=17:52:52.37, $\delta$=+60:23:46.9;
$l=89^\circ.24, b=30^\circ.56$) is one of the JWST calibrator stars
\citep{klc+21}. It is listed by
Simbad\footnote{\url{https://simbad.unistra.fr/simbad/}} as a bright
($V=6.85$) emission-line star of type A2e. The {\it Gaia} parallax
is $5.208\pm 0.04\,$mas. At this distance the proper motion amounts
to $-1.8\,{\rm km\,s^{-1}}$ in right ascension and 39\,km\,s$^{-1}$
along the declination axis. The star with an  estimated age of
310\,Myr is not a part of any moving group.  The radial velocity
is $-16\,{\rm km\,s^{-1}}$. \hd\  was a part of a sample of A stars
with debris that was observed with the Spitzer Space Telescope
\citep{srs+06}.

Given the newness of the MIRI data we checked its calibration in a
number of ways. First to assess the point source photometry we
extracted the signal at 12.814 and 15.5505 \mum\ in a 3\arcsec-radius
and compared those with observed by the WISE satellite \citep{Wright2010}
and extrapolated using a Rayleigh-Jeans law, $F_\nu\propto \nu^{2}$.
The agreement is excellent: WISE 0.0676$\pm$0.001 Jy and  0.0459$\pm$0.001
at 12.814 and 15.5505 \mum, respectively vs. MIRI 0.0674$\pm$0.001
Jy and 0.0410$\pm$0.001 Jy  with an assumed calibration uncertainty
of 2\%. These correspond to differences of 2 and 10\% at the two
wavelengths.

\section{NGC\,6543}
 \label{sec:NGC6543}
 
The planetary nebula NGC\, 6543 was observed during the commissioning
of MIRI-MRS (PID\#1047 and \#1031).  The log can be found in
Table~\ref{tab:NGC6543_log}.

\begin{deluxetable*}{lrrrrr}[hbt]

\tablecaption{Summary of log of data sets for NGC\, 6543
 \label{tab:NGC6543_log}}
\tablehead{
\colhead{name}&
\colhead{$\alpha$\,(deg)} &
\colhead{$\delta$\,(deg)} &
\colhead{series} &\colhead{Aperture (\arcsec)}&
\colhead{$\tau$(s)}
}
\startdata
NGC6543 ([NeII],[NeIII])  & 269.63667 &66.63133 & \texttt{jw01047-o001\_t006}  &4.0 & 4129\\
NGC6543 ([ArII])  &269.64313 & 66.63344 & \texttt{jw01031-c1003\_t010}  &0.7 &4129 \\
NGC6543 ([SIII])  &269.64313 & 66.63344 & \texttt{jw01031-o012\_t010}  &2.0 & 4129 \\
\enddata
 \tablecomments{ The ``name" refers to our assigned name for the
 data set. The next two columns are J2000 right ascension and
 declination, followed by file name identifier  of the Level-3
 MIRI-MRS pipeline data sets. The next column is the radius of the
 photometric aperture followed by the integration time.  }
\end{deluxetable*}

%
%
%

\begin{figure*}[htbp]
\centering
 \includegraphics[width=0.4\textwidth]{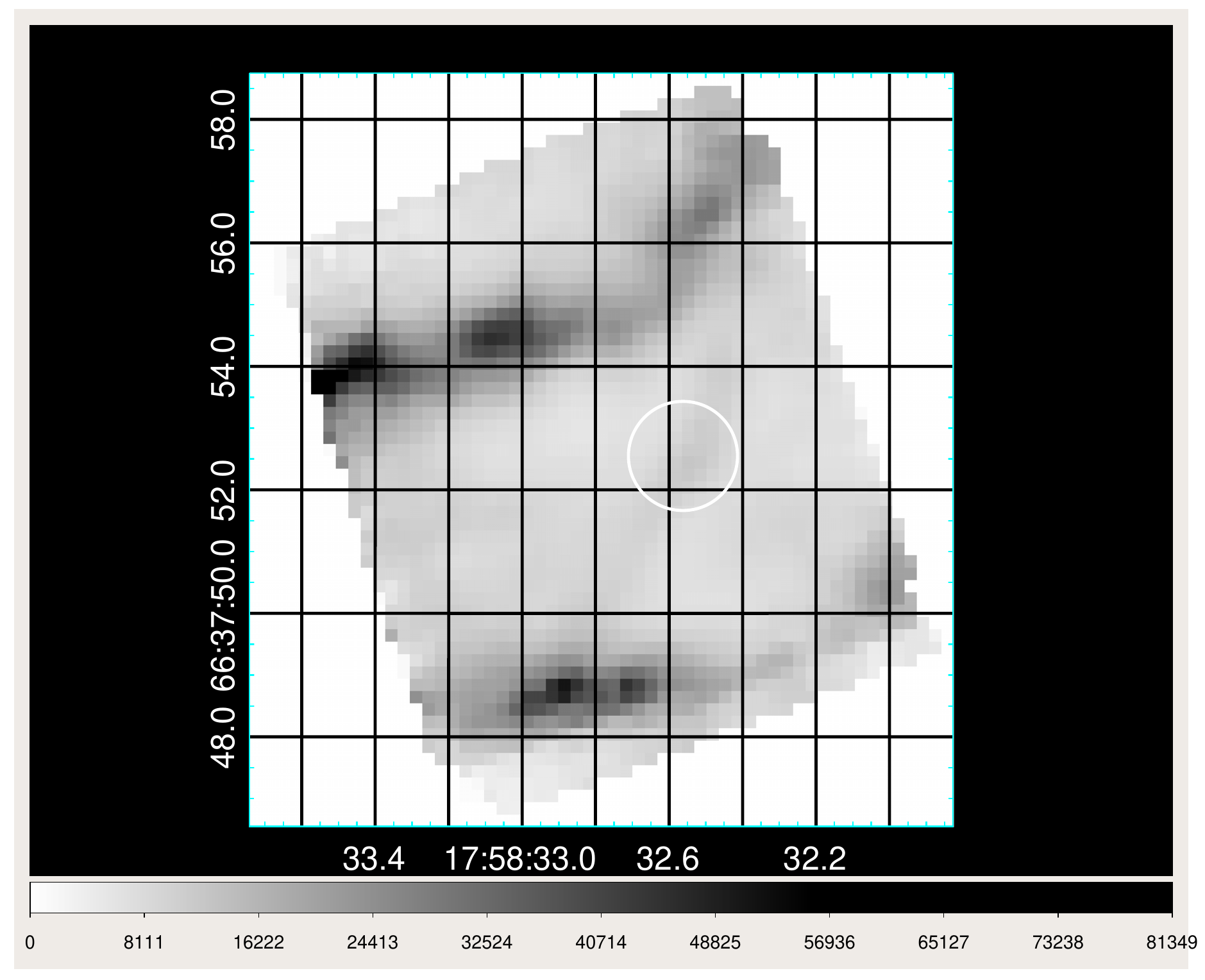}\qquad
  \includegraphics[width=0.4\textwidth]{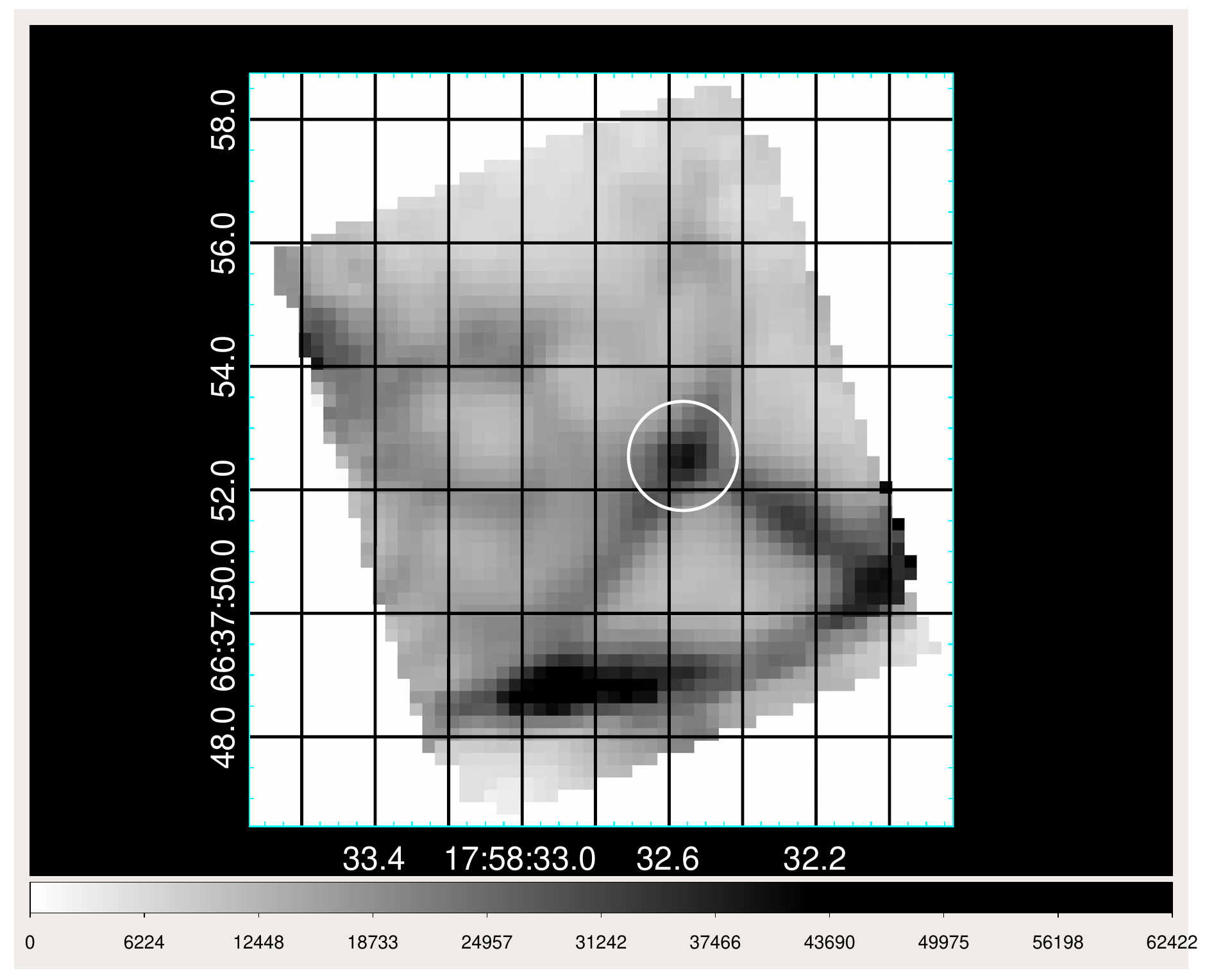}\\
 \includegraphics[width=0.4\textwidth]{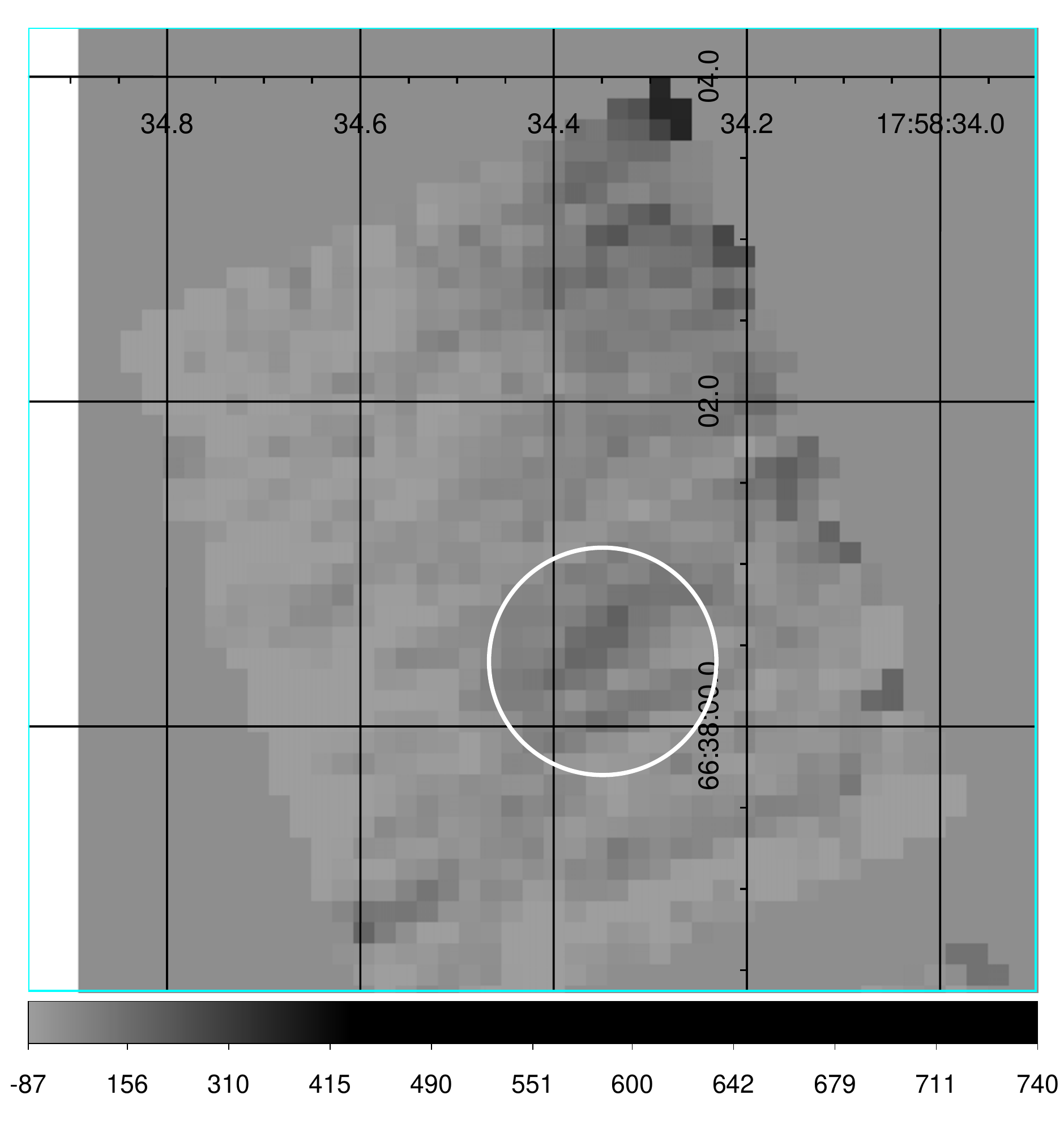}\qquad
\includegraphics[width=0.4\textwidth]{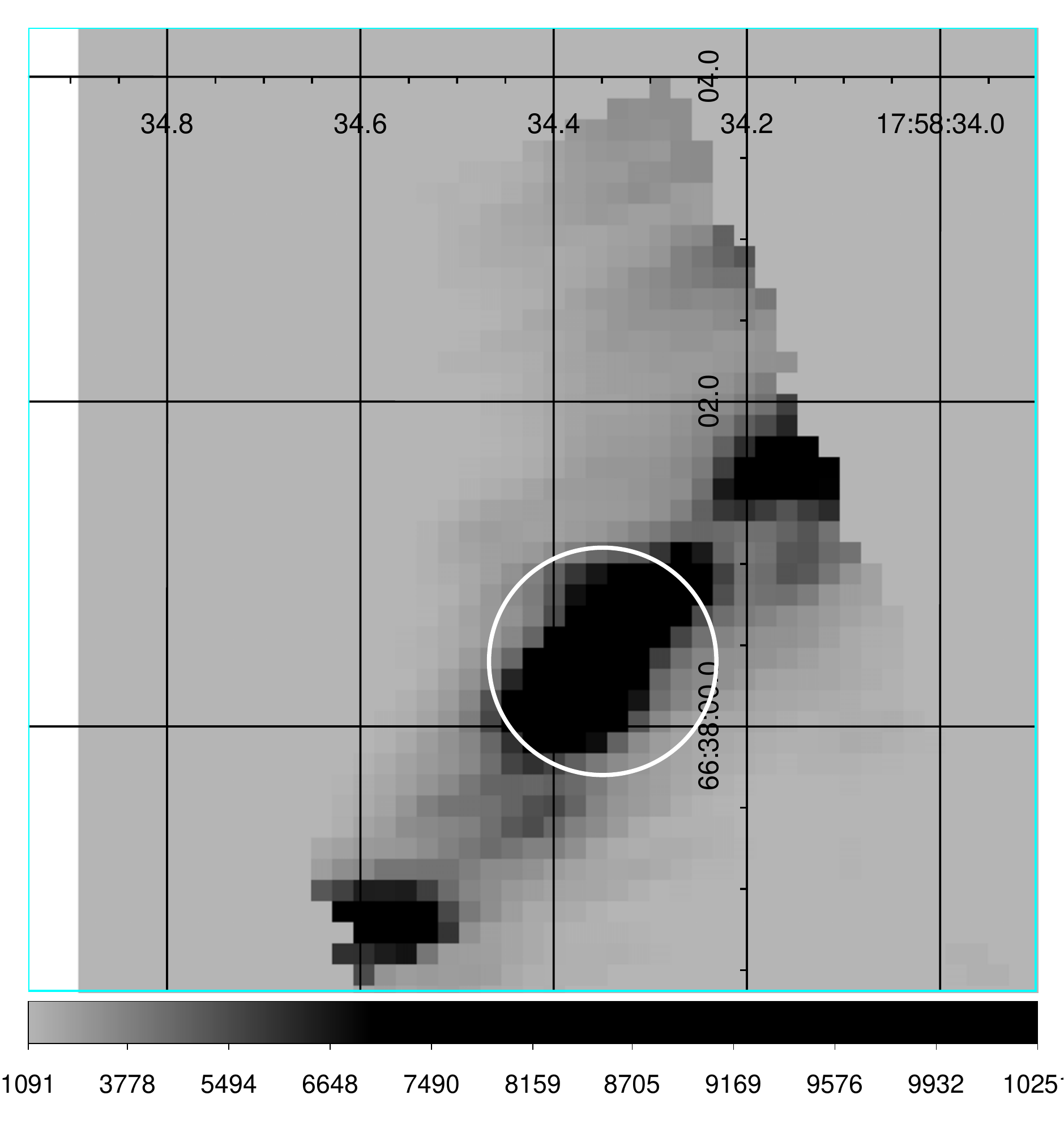}\\
 \caption{(Top) Location of aperture on knot of [NeII] emission in
  the Planetary Nebula NGC6543 off and nn the 12.8 \mum\ line (right
  and left, respectively.  (Bottom ) Emission from the knot in the
  6.95 \mum\ of [ArII] off and on the line }
 \label{fig:NGC6543Image}
\end{figure*}

 \begin{figure*}[htbp]
  \includegraphics[width=0.45\textwidth]{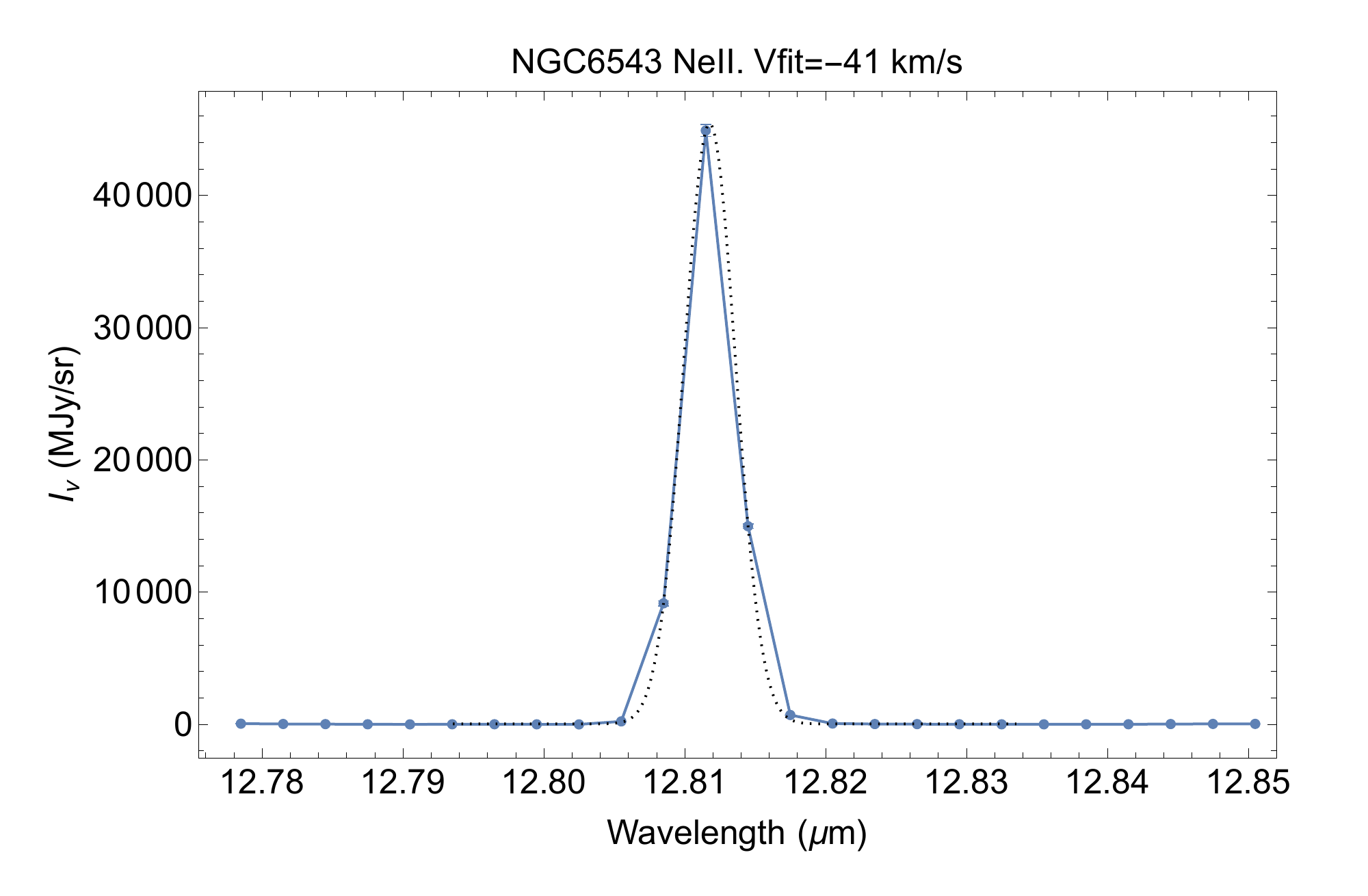}\includegraphics[width=0.45\textwidth]{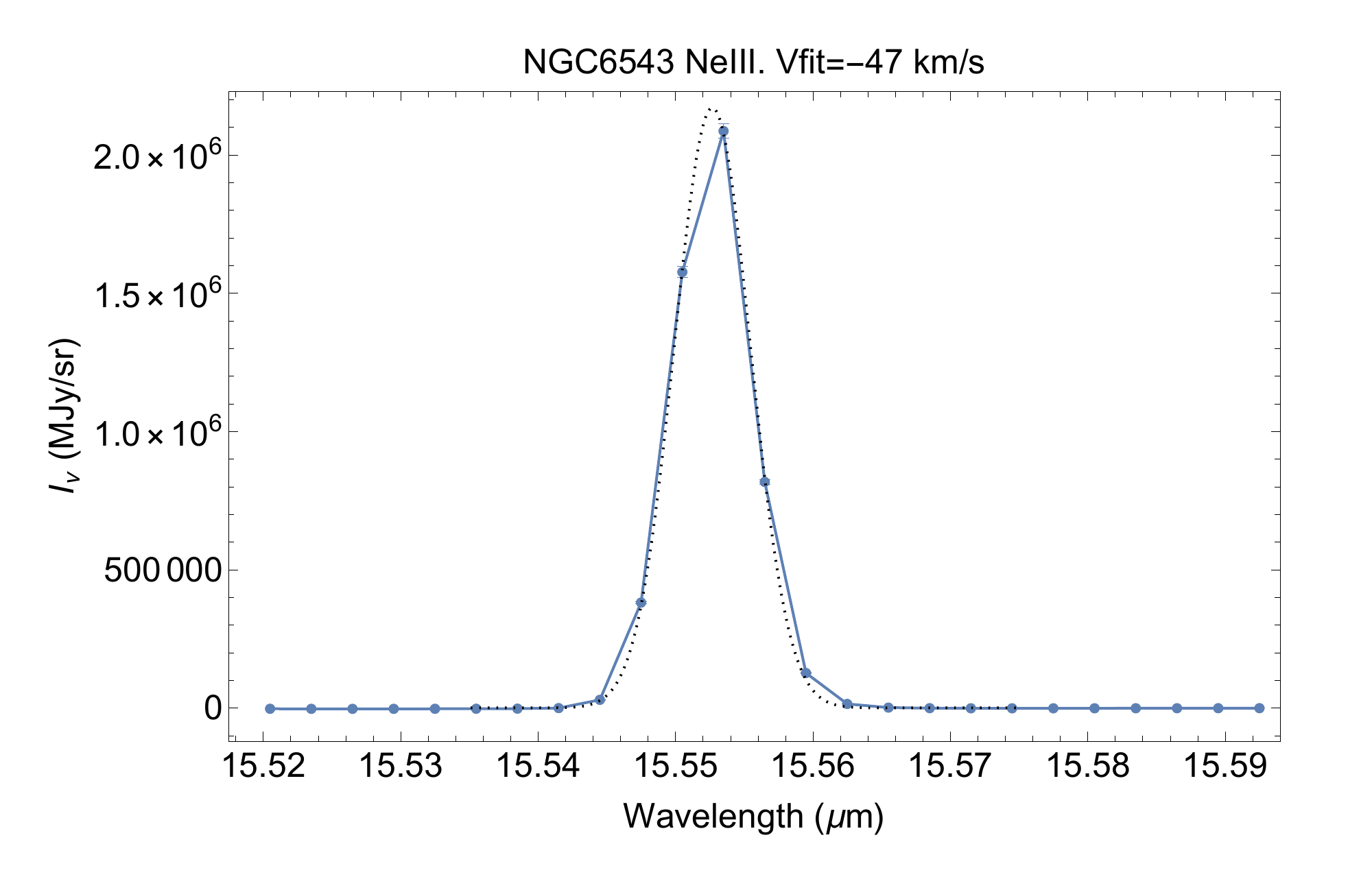}\\
 \includegraphics[width=0.45\textwidth]{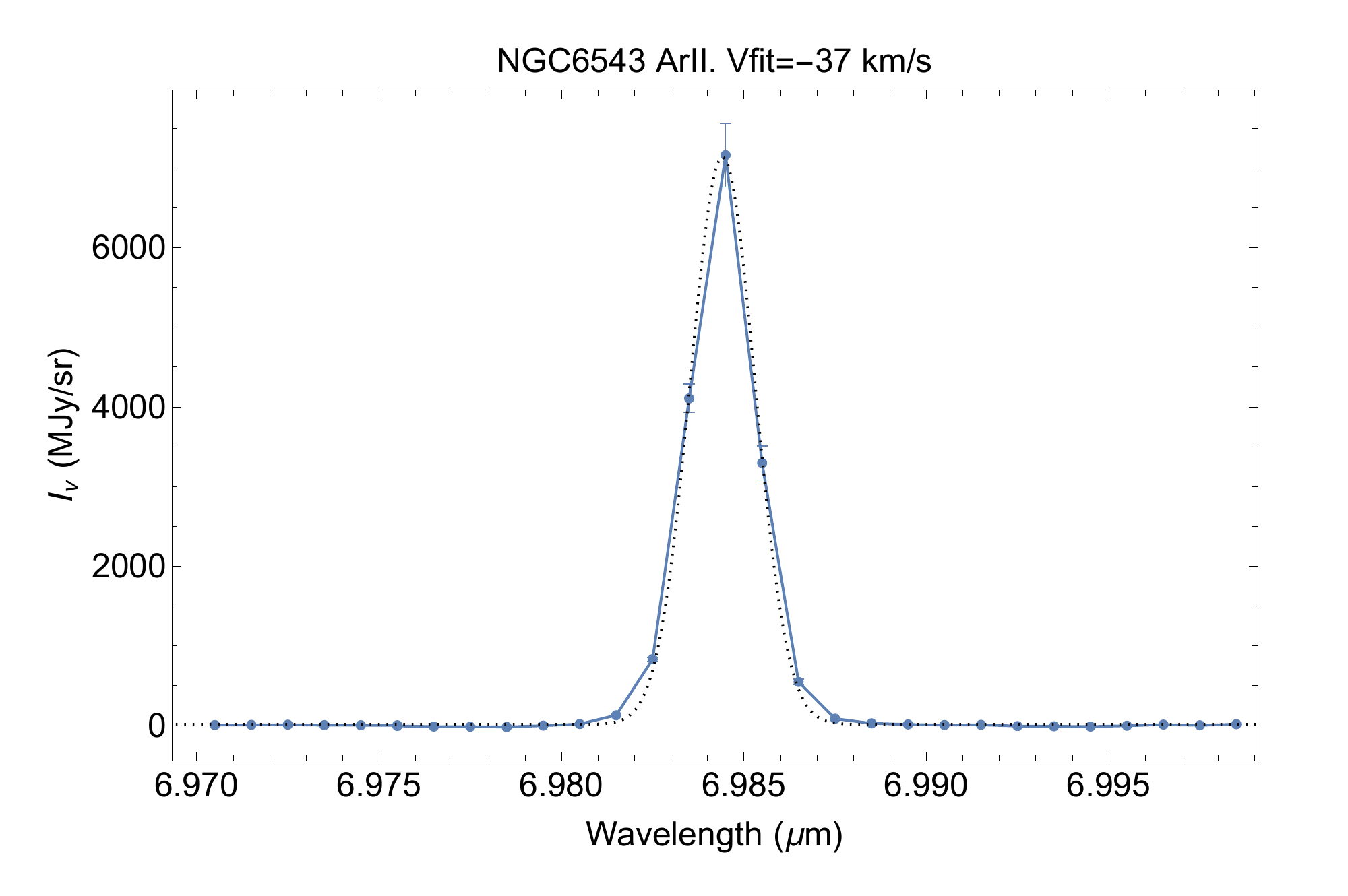}\includegraphics[width=0.45\textwidth]{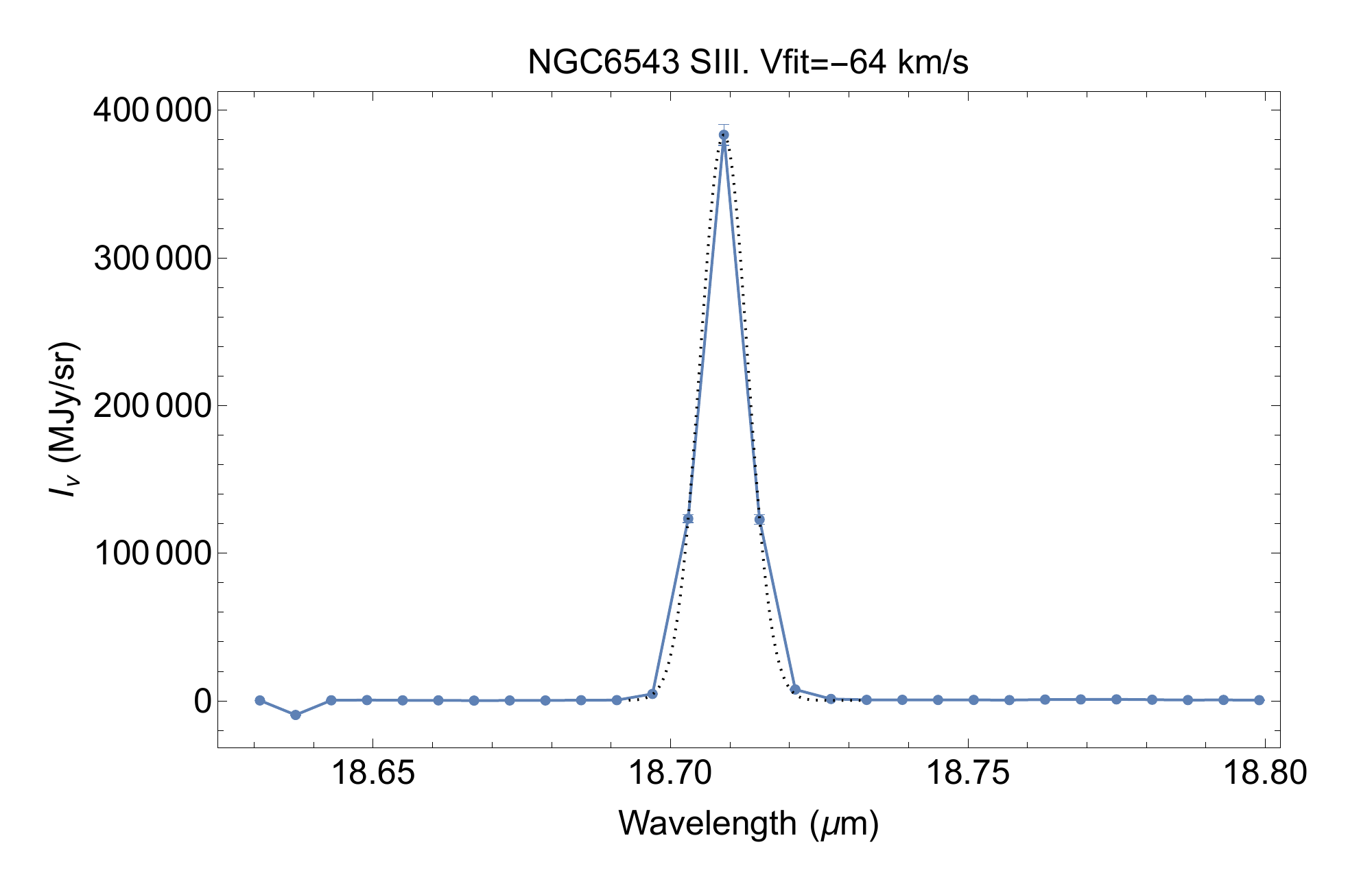}\\
  \caption{top) Emission from the knot in the 12.8 \mum\ (right)
  and 15.55 \mum\ (left) lines of [NeII] and [NeIII], respectively.
  bottom) emission in [ArII] and [SIII]. In all cases the emission
  lines are unresolved. The dotted black lines denote the results
  of  Gaussian fits to the emission with the FHWMs at leach wavelength
  used in the analysis of the \hd\ observations. }
 \label{fig:NGC6543Spectra}
\end{figure*}

\begin{deluxetable*}{llllllr}[htb]
 \centering
 \tablecaption{Analysis of NGC\,6543\label{tab:datafits}}
\tablehead{
\colhead{} &\colhead{}  & \colhead{Ampl} & \colhead{$\lambda$} &\colhead{FWHM} &   \colhead{V$_{\rm fit}$}&\colhead{Integrated Inten.}\\
\colhead{Object} &\colhead{Line}  &
 \colhead{MJy $sr^{-1}$} & 
 \colhead{$\mu$m} &
 \colhead{$\mu$m} &   
 \colhead{km\,s$^{-1}$}&
 \colhead{${\rm erg\,cm^{-2}\,s\,^{-1}\, arcsec^{-2}}$}
}
\startdata
NGC\, 6543&[ArII]&(7.120$\pm0.045)\times10^4$&6.98441&0.00208&$-36.9$&
($2.36\pm0.01)\times 10^{-14}$\\
NGC\, 6543&[NeII]&(4.539$\pm0.008\times10^4$&12.8118&0.00431&$-41.50$&
($8.976\pm0.001)\times 10^{-14}$\\
NGC\, 6543&[NeIII]&(2.17$\pm0.008)\times10^{6}$&15.5527&0.00649&$-46.63$&
$(4.37\pm0.002)\times10^{-12}$\\ 
NGC\, 6543&[SIII]&(3.83$\pm0.002)\times10^{5}$&18.709&0.00938&$-64\pm8^1$&
($7.716\pm0.004)\times 10^{-13}$\\ 
\enddata
\tablecomments{\small $^1$Velocity uncertainty due to uncertainty in rest wavelength.}
\end{deluxetable*}

Figure~\ref{fig:NGC6543Image} shows knots  of [NeII] and [ArII]
emission taken in slightly different regions of the nebula and
measured in and out of the lines.  The resultant line intensities
are so strong that the data filtering required for the weak lines
seen toward \hd\ was  not required. A linear baseline was subtracted
from the median values within the given apertures
(Table~\ref{tab:NGC6543_log}) as shown in Figure~\ref{fig:NGC6543Spectra}.
These data demonstrate the ability of MIRI-MRS  to detect  robustly
the lines of  [NeII], [NeIII], [ArII] and [SIII]. Table~\ref{tab:datafits}
summarizes the observational data and derived parameters. The
intrinsic widths of lines within NGC\, 6543 are 10 to 15$\,{\rm
km\,s^{-1}}$ \citep{bmw+92} and are thus unresolved at the $R\sim2700$
spectral resolution of MRS. We fitted the line profiles with a
Gaussian\footnote{We used the $Mathematica$ function $NonlinearModelFit$
to generate the fits to the spectral features and to calculate the
derived parameters.} and report the full width at half-maximum
(FWHM) in Table~\ref{tab:datafits}.  As described in \citet{wpg+15}, the MRS spectra are under
sampled which accounts for the narrowness of the spectra presented
in Figure~\ref{fig:NGC6543Spectra}.

Table~\ref{tab:datafits}  gives  resultant heliocentric velocities
for the different lines ranging from $-40$ to $-68\,{\rm km\,s^{-1}}$.
Although the  systemic velocity of NGC\, 6543 is about $-68\,{\rm
km\,s^{-1}}$, observations of [OIII] at 5007$\AA$ averaged over
slit lengths of 66--129\arcsec \citep{bmw+92} showed  significant
velocity structure from one knot to another suggesting  internal
motions of a few tens of ${\rm km\,s^{-1}}$. In particular, the
region called ``A3"  \citep{bmw+92} close to the site of the MRS
observations, shows velocities as low as $-50\,{\rm km\, s^{-1}}$
\citep{mitchell+05}.  Finally, we compared MIRI spectra of NGC6543
in the NeII and NeIII lines obtained with the Spitzer IRS spectrometer.
While the spectral resolution is quite different ($\mathcal{R}$=600
vs 2700) and the positions in the nebula different by up to an
arc-minute, the agreement between the two spectra is within a factor
of two.

\end{document}